\documentclass[a4paper,fleqn,usenatbib]{mnras}
\usepackage{newtxtext,newtxmath}

\usepackage[T1]{fontenc}
\usepackage{ae,aecompl}

\usepackage{amsmath}	
\usepackage{amssymb}	
\usepackage{lscape}
\usepackage{graphicx}
\usepackage{bm}

\title[Ionization degree in the primordial clouds]
{Ionization degree and magnetic diffusivity in the primordial star-forming clouds}

\author[D. Nakauchi, K. Omukai, and H. Susa]{
Daisuke Nakauchi$^{1}$\thanks{E-mail: nakauchi@astr.tohoku.ac.jp},
Kazuyuki Omukai$^{1}$, and Hajime Susa$^{2}$
\\
$^{1}$Astronomical Institute, Tohoku University, Aoba, Sendai 980-8578, Japan \\
$^{2}$Department of Physics, Konan University, Higashi-Nada, Kobe, 658-0072, Japan
}
\date{Accepted XXX. Received YYY; in original form ZZZ}

\pubyear{2019}

\begin{document}
\label{firstpage}
\pagerange{\pageref{firstpage}--\pageref{lastpage}}
\maketitle

\begin{abstract}
Magnetic fields play such roles in star formation 
as the angular momentum transport in star-forming clouds, thereby 
controlling circumstellar disc formation and even binary star formation efficiency. 
The coupling between the magnetic field and gas is determined by the ionization degree in the gas.
Here, we calculate the thermal and chemical evolution of the primordial gas
by solving chemical reaction network where all the reactions are reversed.
We find that at $\sim 10^{14}\mbox{-}10^{18}\ {\rm cm}^{-3}$, the ionization degree becomes 100-1000 times higher 
than the previous results due to the lithium ionization by thermal photons trapped in the cloud, which has been omitted so far.
We construct the minimal chemical network which can reproduce correctly the ionization degree 
as well as the thermal evolution by extracting 36 reactions among 13 species.   
Using the obtained ionization degree, we evaluate the magnetic field diffusivity.   
We find that the field dissipation can be neglected for global fields coherent over 
$\gtrsim$ a tenth of the cloud size as long as the field is not so strong as to prohibit the collapse.
With magnetic fields strong enough for ambipolar diffusion heating to be significant,
the magnetic pressure effects to slow down the collapse and to reduce the compressional heating
become more important, and the temperature actually becomes lower than in the no-field case. 
\end{abstract}



\section{Introduction}\label{sec:intro}

Population III~(Pop III) stars are believed to have played pivotal roles in the cosmic structure formation.  
Radiation emitted by them ended the dark age of the Universe 
and controlled the subsequent star-formation by such feedback as ionization and heating~\citep[e.g.,][]{Ciardi2005,Bromm2011}.
Their supernova~(SN) explosions enrich the interstellar medium~(ISM) with metals and triggers the formation of metal-poor stars, some of which are still observed in the Galactic halo~\citep{Ritter2012,Chiaki2018}.
The extent of such feedback depends on the Pop III stellar evolution, which is basically set by the mass and possibly also by the spin.

Pop III stars can be a source of gravitational waves, if they are formed as a binary system and merge within the Hubble time.
In fact, some authors suggested that massive binary black holes~(BHs) of $\sim 30\ {\rm M}_\odot$, which were found by the advanced LIGO and Virgo,
can be originated from Pop III binary systems~\citep[e.g.,][]{Kinugawa2014, Kinugawa2016}.
The merger rate of binary BHs depends on the formation efficiency and nature of Pop III binaries.

According to the recent numerical simulations coupling the multi-dimensional radiation hydrodynamics with the stellar evolution calculation,
the Pop III stars are predicted to be typically very massive, reflecting the high temperature~($\sim$ several 100\ K) in the primordial gas, whose cooling relies solely on 
the rather inefficient H$_2$ cooling due to the lack of metals and dust grains.
Their mass distributes in a wide range from a few 10 to as massive as $1000\ {\rm M}_{\sun}$~\citep{Hosokawa2011, Hosokawa2016, Stacy2012, Stacy2016, Hirano2014, Susa2014}.
Pop III stars can be formed in a binary and multiple-star system, if circumstellar discs are formed and fragment~\citep{Machida_Omukai2008,Stacy2010,Clark2011,Greif2012}.
Also Pop III stars may rotate near the break-up speed~\citep{Stacy2011,Stacy2013},
although the spin of Pop III stars depends on the efficiency of binary formation and still largely uncertain. 

Those conclusions might be altered if magnetic fields are considered.
Magnetic fields affect various aspects of star formation. 
For example, magnetic braking makes the rotation of star-forming clouds slow, thereby controlling 
the formation of a circumstellar disc and a binary system via disc fragmentation~\citep[e.g.,][]{Tsukamoto2016}.
\cite{Machida2013} found by the three-dimensional~(3D) magnetohydrodynamical~(MHD) calculations 
that magnetic braking extracts a large amount of the angular momentum from the clouds and
suppresses the formation of binary and multiple-star systems for the fields stronger 
than $B \sim 10^{-12}\ {\rm G}$ (at $\sim 1\ {\rm cm}^{-3}$).
Similarly, \cite{Hirano_Bromm2018} indicated from a semi-analytical model that
the Pop III stars become slow-rotators for strong fields of $B \gtrsim 10^{-8}\ {\rm G}$~(at $\sim 1\ {\rm cm}^{-3}$).
Magnetic fields can also drive outflows, ejecting a part of the materials in the cloud back to the interstellar space, 
and reduce the star-formation efficiency. 
MHD outflows eject $\sim$ 10\% of the materials from the cloud for $B \gtrsim 10^{-11}\ {\rm G}$~(at $\sim 1\ {\rm cm}^{-3}$)~\citep{Machida2006,Machida2008}.

Various scenarios, from cosmological to astrophysical ones, have been invoked as the origin of the primordial magnetic fields~\citep[e.g.,][]{Ando2010,Widrow2012,Subramanian2016}.
Although the primordial fields may have been orders of magnitude weaker than those in the Galactic ISM~\citep[$\sim 5\ \mu{\rm G}$;][]{Crutcher2010}, 
they can be amplified via the small scale dynamo in a highly turbulent medium~\citep{Sur2010,Sur2012,Federrath2011,Turk2012}.  
In fact, the ISM in the first galaxies is predicted to be highly turbulent as a result of the accretion flows 
according to the cosmological simulations~\citep{Wise2007,Wise2008, Greif2008}. 
Analytical estimate indicates that the field can reach as strong as the equipartition value 
of $\sim 1\ \mu{\rm G}$ at $\sim 1\ {\rm cm}^{-3}$ \citep{Schleicher2010, Schober2012a}.

Magnetic fields can dissipate via ambipolar diffusion and/or Ohmic loss during the pre-stellar collapse since the ionization degree is very low~($\lesssim 10^{-10}$)~\citep{Nakano1986}.
The magnetic diffusivity depends on the ionization degree in the gas. 
By solving the primordial-gas chemical network taking also into account the lithium-related reactions, 
\cite{Maki2004, Maki2007} found that the ionization degree remains high enough to prevent 
the magnetic field dissipation throughout the pre-stellar collapse thanks to the ionization of lithium, which has a small ionization potential. 
Their chemical network is, however, unsatisfactory in the sense that it is not reversed, i.e., includes only a portion of the reverse reactions. 
At high enough densities, the chemical abundances reach the equilibrium values given by the Saha-like equations as a result of the balance between the forward and reverse reactions.  
In the \cite{Maki2004, Maki2007} calculation, the chemical abundance is switched to the equilibrium value by hand at an arbitrarily chosen density $\sim 10^{18}\ {\rm cm}^{-3}$, where the abundances of some chemical species 
become discontinuous.
This signals the incorrect results for the ionization degree around this density, where the magnetic field dissipates most easily.

In this paper, by constructing chemical network where all the reverse reactions are included, 
we calculate the thermal and chemical evolution of the primordial gas during the pre-stellar collapse.
This assures all the chemical abundances smoothly reach the correct chemical equilibrium values at high enough densities. 
The obtained abundances are then used to evaluate the magnetic diffusivities.
We find that the ionization degree is more than two orders of magnitude higher than the previous result
at $\sim 10^{14}-10^{18}\ {\rm cm}^{-3}$ because of the lithium ionization by the thermal photons trapped in the cloud, which was omitted in the previous network.
The higher ionization degree results in the stronger coupling between the gas and magnetic field, suppressing magnetic dissipation up to scales much smaller than in the previous network.

The rest of the paper is organized as follows.
In Section \ref{sec:method}, we describe our model for the cloud dynamics, cooling and heating processes, and chemical network.
The results for thermal evolution and ionization degree are presented in Section \ref{sec:thermal_evolution}.
In Section \ref{sec:magnetic}, after calculating the magnetic diffusivities, we examine the conditions for magnetic  dissipation.
We also present the cases with strong fields, where thermal evolution is affected by the magnetic pressure and magnetic dissipation heating.  
In section \ref{sec:summary}, after briefly summarizing the results, we discuss the uncertainties and implications of our results.
For the cosmological parameters, we adopt the following values: $\Omega_\Lambda = 0.6911, \Omega_{\rm m} = 0.3089, \Omega_{\rm b} = 0.0486$, $H_0 = 67.74\ {\rm km}\ {\rm s}^{-1}\ {\rm Mpc}^{-1}$, and $\rho_{\rm cr, 0} = 8.62 \times 10^{-30}\ {\rm g}\ {\rm cm}^{-3}$~\citep{Planck2016}.

\section{Method}
\label{sec:method}

We consider the gravitational collapse of a spherical symmetric cloud neglecting rotation and turbulence for simplicity. 
Also back-reaction of magnetic fields, magnetic pressure and heating due to magnetic diffusion, is neglected assuming the field is sufficiently weak, 
except in Section \ref{subsec:B_backreact}.
Such a cloud experiences the so-called runaway collapse, i.e., the central part or `core', where the density is highest, 
collapses fastest, leaving the lower density `envelope' almost unevolved~\citep{Larson1969,Penston1969}. 
During the runaway collapse, the self-similar density distribution develops, and the central core with the local Jeans size 
\begin{equation}
\lambda_{\rm J} = \sqrt{\frac{\pi k_{\rm B} T}{G \mu m_{\rm H} \rho}}
\label{eq:lambda_jeans}
\end{equation}
has a uniform density, while the surrounding envelope has a power-law profile with radius.

We calculate the evolution of physical quantities in the central core by using 
a one-zone model~\citep{Omukai2000,Omukai2001,Omukai2012}.
The gas density at the center $\rho$ increases in the local free-fall time:
\begin{equation}
\frac{d \rho}{dt} = \frac{\rho}{t_{\rm col}},
\label{eq:free_fall}
\end{equation}
where 
\begin{equation}
t_{\rm col} =  \sqrt{\frac{ 3 \pi }{32 G \rho_{\rm m}}},
\label{eq:t_str}
\end{equation}
using the total mass density $\rho_{\rm m}$, which consists of the dark matter and gas~($\rho_{\rm m}=\rho_{\rm DM} + \rho$).

The DM density is given by the top-hat density evolution model: 
\begin{equation}
\rho_{\rm DM} = \frac{9}{2} \pi^2 \left(\frac{1 + z_{\rm turn}}{1- \cos \theta}\right)^{3} \rho_{\rm cr, 0} \Omega_{\rm DM},
\label{eq:rho_m}
\end{equation}
where $z_{\rm turn}$ is the turnaround redshift and the parameter $\theta$ is related with redshift $z$ as
\begin{equation}
1+z = (1 + z_{\rm turn}) \left(\frac{\theta - \sin \theta}{\pi}\right)^{-2/3}.
\label{eq:z_rs}
\end{equation}
We start the calculation from the turnaround redshift of $z_{\rm turn} = 16.5$ and $\theta = \pi$, 
when the number density and temperature of the gas are $n_{\rm H, ini} = 5.6 \times 10^{-3}\ {\rm cm}^{-3}$ and $T_{\rm ini} = 22\ {\rm K}$, respectively.
Once the DM density reaches the virialization value $\rho_{\rm DM, vir} = 8 \rho_{\rm DM}(z_{\rm turn})$, we set it constant thereafter.
Note that for the gas density higher than the DM density, i.e., $n_{\rm H} \gtrsim 0.1\ {\rm cm^{-3}}$, the gravity is dominated by the gas and 
the presence of DM essentially does not affect the cloud evolution at all. 

The temperature evolution is followed by solving the energy equation:
\begin{equation}
\frac{de}{dt} = - P \frac{d}{dt}\left(\frac{1}{\rho}\right) - \Lambda_{\rm net},
\label{eq:EoE}
\end{equation}
where the pressure 
\begin{equation}
P = \frac{\rho k_{\rm B} T}{\mu m_{\rm H}}
\label{eq:press}
\end{equation}
the specific internal energy
\begin{equation}
e = \frac{1}{\gamma-1} \frac{P}{\rho},
\label{eq:ene}
\end{equation}
and $\Lambda_{\rm net}$ is the net cooling rate per unit mass. 
The other symbols have their usual meanings.
The net cooling rate $\Lambda_{\rm net}$ consists of the following terms:
\begin{equation}
\Lambda_{\rm net}=\Lambda_{\rm line}+\Lambda_{\rm cont}+\Lambda_{\rm chem}, 
\end{equation}
where $\Lambda_{\rm line}$ is the line cooling due to H Ly$\alpha$, H$_2$, and HD, 
$\Lambda_{\rm cont}$ is the continuum cooling dominantly by H$_2$ collision-induced emission~(CIE), 
and $\Lambda_{\rm chem}$ is the cooling/heating associated with the chemical reactions.
We use the formulation of \cite{Omukai2012} for the cooling/heating processes, but with the following updates.
For H$_2$ cooling, we use the fitting function by \cite{Glover2015b}, considering the transitions due to the collisions between  
H$_2$-H, H$_2$-H$_2$, H$_2$-He \citep{Glover2008}, H$_2$-H$^+$ and H$_2$-$e$ \citep{Glover2015a}.
The photon trapping effect in the optically thick regime is taken into account by using the cooling suppression factor by \cite{Fukushima2018} 
which depends on the column density of the cloud core: $N_{\rm H} = n_{\rm H} \lambda_{\rm J}$.
The HD cooling rate is taken from  \cite{Lipovka2005}, considering the HD-H collisional transition. 

For the elemental abundances of He, D, and Li, we adopt the standard Big Bang nucleosynthesis values with the baryon to photon ratio by PLANCK, $y_{\rm He} = 8.3 \times 10^{-2}, y_{\rm D} = 2.6 \times 10^{-5}$, and $y_{\rm Li} = 4.7 \times 10^{-10}$~\citep{Cyburt2016}.
The initial chemical fractions are assumed to be the typical intergalactic values in the post-recombination era~\citep{Galli2013}, 
$y({{\rm H}^+}) = 10^{-4}$, $y({{\rm H}_2}) = 6 \times 10^{-7}$, and $y({\rm HD}) = 4 \times 10^{-10}$, with the remaining H and D in the neutral atomic state.
The helium is neutral, while the lithium is fully ionized initially.

\begin{table}
\caption{Chemical reactions in our calculation.}
\begin{center}
{\begin{tabular}{llc}
\hline
Number & Reaction & Reference \\
\hline
H1 & $  {\rm H}^+    +   e     \rightleftharpoons   {\rm H}     +   \gamma $ & 1 \\
H2 & $  {\rm H}     +   e     \rightleftharpoons   {\rm H}^-   +   \gamma           $ & 1 \\
H3  & $  {\rm H}^-   +   {\rm H}     \rightleftharpoons   {\rm H}_2   +   e  $ & 2 \\
H4 & $  {\rm H}     +   {\rm H}^+    \rightleftharpoons   {\rm H}_2^+  +   \gamma $ & 3 \\
H5 & $  {\rm H}_2^+  +   {\rm H}     \rightleftharpoons   {\rm H}_2   +   {\rm H}^+   $ & 1 \\
H6 & $3 {\rm H}               \rightleftharpoons   {\rm H}_2   +   {\rm H}       $ & 4 \\
H7 & $2 {\rm H}_2             \rightleftharpoons 2 {\rm H}     +   {\rm H}_2     $ & 1 \\
H8 & $  {\rm H}_2^+  +   {\rm H}_2   \rightleftharpoons   {\rm H}_3^+  +   {\rm H}  $ & 5 \\
H9 & $  {\rm H}_3^+  +   e     \rightleftharpoons   {\rm H}_2   +   {\rm H}   $ & 5 \\
H10 & $  {\rm H}_3^+  +   e     \rightleftharpoons 3 {\rm H}  $ & 5 \\
H11 & $  {\rm H}^-   +   {\rm H}^+    \rightleftharpoons 2 {\rm H}                $ & 1 \\
H12 & $  {\rm H}_2^+  +   e     \rightleftharpoons 2 {\rm H}                $ & 1 \\
H13 & $  {\rm H}^-   +   {\rm H}     \rightleftharpoons 2 {\rm H}     +   e   $ & 1 \\
H14 & $  {\rm H}     +   {\rm H}     \rightleftharpoons  {\rm H}    +    e    +   {\rm H}^+$ & 6 \\ 
H15 & $  {\rm H}     +   e     \rightleftharpoons   {\rm H}^+    + 2 e  $ & 1 \\ 
H16 & $  {\rm H}_2   +   e     \rightleftharpoons    2 {\rm H}     +   e             $ & 1 \\
H17 & $  {\rm H}^-   +   e     \rightleftharpoons   {\rm H}     + 2 e             $ & 1 \\
H18 & $  {\rm H}^-   +   {\rm H}^+    \rightleftharpoons   {\rm H}_2^+  +   e          $ & 1 \\
H19 & $  {\rm H}_2^+  +   {\rm H}^-   \rightleftharpoons   {\rm H}_2   +   {\rm H}       $ & 1 \\  
H20 & $  {\rm H}_2^+  +   {\rm H}^-   \rightleftharpoons   {\rm H}_3^+   +  e       $ & 5 \\  
H21 & $  {\rm H}      +    {\rm H}_2^+   \rightleftharpoons   {\rm H}_3^+   +   \gamma       $ & 5 \\  
H22 & $  {\rm H}_2  +   {\rm H}^+   \rightleftharpoons   {\rm H}_3^+   +   \gamma       $ & 7 \\
H23 & $  {\rm H}_3^+  +   {\rm H}^-   \rightleftharpoons 2 {\rm H}_2  $ & 5 \\
H24 & $  {\rm H}^-   +   {\rm H}_2^+  \rightleftharpoons 3 {\rm H}    $ & 1 \\
He1 & $  {\rm He}^+  +   e     \rightleftharpoons   {\rm He}   +   \gamma       $ & 1 \\ 
He2 & $  {\rm He}^+  +   {\rm H}_2   \rightleftharpoons   {\rm H}^+    +   {\rm H}     +   {\rm He} $ & 1 \\
He3 & $  {\rm He}   +   e     \rightleftharpoons   {\rm He}^+  + 2 e    $ & 1 \\
He4 & $  {\rm He}^+  +   e     \rightleftharpoons   {\rm He}^{++} + 2 e $ & 1 \\
He5 & $  {\rm He}^{++} +   e     \rightleftharpoons   {\rm He}^+  +   \gamma        $ & 1 \\
He6 & $  {\rm He}^+  +   {\rm H}     \rightleftharpoons   {\rm H}^+    +   {\rm He}      $ & 1 \\
He7 & $  {\rm He}^+  +   {\rm H}^-   \rightleftharpoons   {\rm H}     +   {\rm He}     $ & 1 \\
He8 & $  {\rm He}^+  +   {\rm H}_2   \rightleftharpoons   {\rm H}_2^+  +   {\rm He} $ & 1 \\
He9 & $  {\rm H}_2   +   {\rm He}   \rightleftharpoons 2 {\rm H}     +   {\rm He}     $ & 1 \\
He10 & $  {\rm He}   +   {\rm H}^-   \rightleftharpoons   {\rm He}   +   {\rm H}   +   e  $ & 1 \\
He11 & $  {\rm H}^+  +   {\rm He}   \rightleftharpoons   {\rm HeH}^+  +  \gamma    $ & 5 \\
He12 & $  {\rm H}_2^+  +   {\rm He}   \rightleftharpoons   {\rm HeH}^+  +   {\rm H}    $ & 5 \\
He13 & $ {\rm HeH}^+  +   {\rm H}_2   \rightleftharpoons   {\rm H}_3^+  +   {\rm He} $ & 5 \\
He14 & $  {\rm HeH}^+  +   e     \rightleftharpoons   {\rm H}     +   {\rm He}   $ & 5 \\
D1 & $    {\rm D}  +   {\rm H}^+   \rightleftharpoons   {\rm D}^+  +   {\rm H}    $ & 1 \\
D2 & $    {\rm D}  +   {\rm H}_2  \rightleftharpoons   {\rm H}    +   {\rm HD}     $ & 1 \\
D3 & $    {\rm D}^+ +   {\rm H}_2  \rightleftharpoons   {\rm H}^+   +   {\rm HD}    $ & 1 \\
D4 & $    {\rm HD}  +   e    \rightleftharpoons  {\rm H}     +   {\rm D}   +   e  $ & 1 \\
D5 & $    {\rm HD}  +   {\rm He}  \rightleftharpoons  {\rm H}     +   {\rm D}   +   {\rm He} $ & 1 \\
D6 & $    {\rm HD}  +   {\rm H}_2  \rightleftharpoons  {\rm H}     +   {\rm D}   +   {\rm H}_2  $ & 1 \\
D7 & $    {\rm HD}  +   {\rm H}    \rightleftharpoons 2 {\rm H}    +   {\rm D}      $ & 1 \\
D8 & $    {\rm D}^+ +   e    \rightleftharpoons   {\rm D}   +   \gamma $ & 1 \\ 
D9 & $    {\rm D}  +   {\rm H}    \rightleftharpoons   {\rm HD}  +   \gamma $ & 1 \\
D10 & $    {\rm HD}^+ +   {\rm H}    \rightleftharpoons   {\rm H}^+   +   {\rm HD}    $ & 1 \\
D11 & $    {\rm D}  +   {\rm H}^+   \rightleftharpoons   {\rm HD}^+ +   \gamma  $ & 1 \\
D12 & $    {\rm D}^+ +   {\rm H}    \rightleftharpoons   {\rm HD}^+ +   \gamma    $ & 1 \\
D13 & $     {\rm HD}^+ +   e    \rightleftharpoons   {\rm H}    +   {\rm D} $ & 1 \\
D14 & $     {\rm D}  +   e    \rightleftharpoons   {\rm D}^-  +   \gamma    $ & 1 \\
D15 & $     {\rm D}^+ +   {\rm D}^-  \rightleftharpoons   2{\rm D} $ & 1 \\
D16 & $     {\rm H}^+  +   {\rm D}^-  \rightleftharpoons   {\rm D}   +   {\rm H}  $ & 1 \\
D17 & $     {\rm H}^- +   {\rm D}   \rightleftharpoons   {\rm H}    +   {\rm D}^-    $ & 1 \\
D18 & $     {\rm D}^- +   {\rm H}    \rightleftharpoons   {\rm HD}  +   e      $ & 1 \\
D19 & $  {\rm D}  +   e    \rightleftharpoons   {\rm D}^+  +  2 e  $ & 1 \\
D20 & $  {\rm He}^+ +   {\rm D}   \rightleftharpoons   {\rm D}^+  +   {\rm He}      $ & 1 \\
D21 & $  {\rm H}_2^+ +   {\rm D}   \rightleftharpoons  {\rm HD}^+  +   {\rm H}       $ & 1 \\
D22 & $  {\rm HD}^+ +   {\rm D}   \rightleftharpoons  {\rm HD}   +   {\rm D}^+    $ & 1 \\
D23 & $  {\rm HD} +   e    \rightleftharpoons  {\rm D}    +   {\rm H}^-    $ & 1 \\
D24 & $  {\rm H}^+  +   {\rm D}^-  \rightleftharpoons  {\rm HD}^+  +   e    $ & 1 \\
D25 & $  {\rm D}^+ +   {\rm H}^-  \rightleftharpoons  {\rm HD}^+  +   e    $ & 1 \\
\hline
\end{tabular}}
\end{center}
\label{tab:chem_react}
\end{table}

\begin{table}
\contcaption{}
\begin{center}
{\begin{tabular}{llc}
\hline
Number & Reaction & Reference \\
\hline
D26 & $  {\rm D}^- +   e    \rightleftharpoons  {\rm D}    + 2 e   $ & 1 \\
D27 & $  {\rm D}^- +   {\rm H}    \rightleftharpoons  {\rm D}    +   {\rm H}    +   e  $ & 1 \\
D28 & $  {\rm D}^- +   {\rm He}  \rightleftharpoons  {\rm D}    +   {\rm He}  +   e  $ & 1 \\
D29 & $  {\rm D}^+ +   {\rm H}^-  \rightleftharpoons  {\rm D}    +   {\rm H}    $ & 1 \\
D30 & $  {\rm H}_2^+ +   {\rm D}^-  \rightleftharpoons  {\rm H}_2   +   {\rm D}   $ & 1 \\
D31 & $  {\rm H}_2^+ +   {\rm D}^-  \rightleftharpoons 2 {\rm H}    +   {\rm D}    $ & 1 \\
D32 & $  {\rm HD}^+ +   {\rm H}^-  \rightleftharpoons   {\rm HD}  +   {\rm H}   $ & 1 \\
D33 & $  {\rm HD}^+ +   {\rm H}^-  \rightleftharpoons   {\rm D}   + 2 {\rm H}    $ & 1 \\
D34 & $  {\rm HD}^+ +   {\rm D}^-  \rightleftharpoons  {\rm HD}   +   {\rm D}  $ & 1 \\
D35 & $  {\rm HD}^+ +   {\rm D}^-  \rightleftharpoons  2 {\rm D}  +   {\rm H}    $ & 1 \\
D36 & $  {\rm He}^+ +   {\rm D}^-  \rightleftharpoons  {\rm He}   +   {\rm D}       $ & 1 \\
D37 & $  {\rm D}  +   {\rm H}_2^+ \rightleftharpoons  {\rm H}_2   +   {\rm D}^+    $ & 1 \\
D38 & $ {\rm H}_2^+ +   {\rm D}   \rightleftharpoons  {\rm HD}   +   {\rm H}^+      $ & 1 \\
D39 & $ {\rm HD}^+ +   {\rm H}    \rightleftharpoons  {\rm H}_2   +   {\rm D}^+     $ & 1 \\
D40 & $ {\rm HD}  +   {\rm He}^+ \rightleftharpoons  {\rm HD}^+  +   {\rm He}             $ & 1 \\
D41 & $ {\rm HD}  +   {\rm He}^+ \rightleftharpoons  {\rm He}   +   {\rm H}^+   +   {\rm D}   $ & 1 \\
D42 & $ {\rm HD}  +   {\rm He}^+ \rightleftharpoons  {\rm He}   +   {\rm H}    +   {\rm D}^+    $ & 1 \\
Li1 & $  {\rm Li}^+  +   e    \rightleftharpoons   {\rm Li}   +   \gamma   $ & 8 \\
Li2 & $  {\rm Li}^+  +   {\rm H}^-  \rightleftharpoons   {\rm Li}   +   {\rm H}    $ & 8 \\
Li3 & $  {\rm Li}   +   {\rm H}^+   \rightleftharpoons   {\rm Li}^+  +   {\rm H}   +   \gamma $ & 8 \\
Li4 & $  {\rm LiH}  +   {\rm H}    \rightleftharpoons   {\rm Li}   +   {\rm H}_2    $ & 8 \\
Li5 & $  {\rm Li}   +   {\rm H}    \rightleftharpoons   {\rm LiH}  +   \gamma   $ & 8 \\
Li6 & $  {\rm Li}^-  +   {\rm H}^+   \rightleftharpoons   {\rm Li}   +   {\rm H}   $ & 8 \\
Li7 & $  {\rm Li}   +   e    \rightleftharpoons   {\rm Li}^-  +   \gamma    $ & 8 \\
Li8 & $  {\rm Li}   +   {\rm H}^+   \rightleftharpoons   {\rm Li}^+  +   {\rm H}    $ & 8 \\
Li9 & $  {\rm Li}   +   {\rm H}^-  \rightleftharpoons   {\rm LiH}  +   e   $ & 8 \\
Li10 & $  {\rm Li}^-  +   {\rm H}    \rightleftharpoons   {\rm LiH}  +   e    $ & 8 \\
Li11 & $  {\rm LiH}  +   {\rm H}^+   \rightleftharpoons   {\rm LiH}^+ +   {\rm H}   $ & 8 \\
Li12 & $  {\rm Li}^+  +   {\rm H}    \rightleftharpoons   {\rm LiH}^+ +   \gamma  $ & 8 \\
Li13 & $  {\rm Li}   +   {\rm H}^+   \rightleftharpoons   {\rm LiH}^+ +   \gamma    $ & 8 \\
Li14 & $  {\rm LiH}  +   {\rm H}^+   \rightleftharpoons   {\rm Li}^+  +   {\rm H}_2    $ & 8 \\
Li15 & $  {\rm LiH}^+ +   e    \rightleftharpoons   {\rm Li}   +   {\rm H}   $ & 8 \\
Li16 & $  {\rm LiH}^+ +   {\rm H}    \rightleftharpoons   {\rm Li}   +   {\rm H}_2^+   $ & 8 \\
Li17 & $  {\rm LiH}^+ +   {\rm H}    \rightleftharpoons   {\rm Li}^+  +   {\rm H}_2     $ & 8 \\
Li18 & $  {\rm Li}   +   {\rm H}_2^+ \rightleftharpoons   {\rm LiH}  +   {\rm H}^+   $ & 8 \\
Li19 & $  {\rm Li}^{++} +   e    \rightleftharpoons   {\rm Li}^+  +   \gamma   $ & 9 \\
Li20 & $ {\rm Li}^{+++} +   e    \rightleftharpoons   {\rm Li}^{++} +   \gamma  $ & 9 \\
Li21 & $  {\rm Li}^+  +   {\rm D}^-  \rightleftharpoons  {\rm Li}    +   {\rm D}  $ & 9 \\
Li22 & $  {\rm Li}^-  +   {\rm D}^+  \rightleftharpoons  {\rm Li}    +   {\rm D}   $ & 9 \\
Li23 & $  {\rm Li}   +   e    \rightleftharpoons  {\rm Li}^+   +   2e  $ & 9 \\
Li24 & $  {\rm Li}^+  +   e    \rightleftharpoons  {\rm Li}^{++}  +   2e   $ & 9 \\
Li25 & $  {\rm Li}^{++} +   e    \rightleftharpoons  {\rm Li}^{+++} +   2e   $ & 9 \\
Li26 & $  {\rm Li}   +   2{\rm H}   \rightleftharpoons  {\rm LiH}   +   {\rm H}   $ & 10 \\
Li27 & $  {\rm Li}   +   {\rm H}   +   {\rm H}_2  \rightleftharpoons  {\rm LiH}   +   {\rm H}_2  $ & 10 \\
CR1 & $ {\rm H} + {\rm CR} \rightarrow {\rm H}^+ + e$ & 11 \\
CR2 & $ {\rm H}_2 + {\rm CR} \rightarrow {\rm H}^+ + {\rm H} + e$ & 11  \\
CR3 & $ {\rm H}_2 + {\rm CR} \rightarrow {\rm H}_2^+ + e$ & 11  \\
CR4 & $ {\rm H}_2 + {\rm CR} \rightarrow 2{\rm H}$  & 11  \\
CR5 & $ {\rm H}_2 + {\rm CR} \rightarrow {\rm H}^+ + {\rm H}^-$ & 11  \\
CR6 & $ {\rm He} + {\rm CR} \rightarrow {\rm He}^+ + e$ & 11  \\
CR7 & $ {\rm H} + {\rm CR ph.} \rightarrow {\rm H}^+ + e$ & 11  \\
CR8 & $ {\rm H}^- + {\rm CR ph.} \rightarrow {\rm H} + e$ & 11  \\
CR9 & $ {\rm He} + {\rm CR ph.} \rightarrow {\rm He}^+ + e$ & 11  \\
\hline
\end{tabular}}
\end{center}
{\bf Notes.} The rate coefficients of the reverse reactions are calculated from the principle of detailed balance. \\
{\bf References.} (1) \cite{Glover2008}, (2) \cite{Kreckel2010}, (3) \cite{Coppola2011}, (4) \cite{Forrey2013}, (5) \cite{Glover2009}, (6) \cite{Lenzuni1991}, (7) \cite{Stancil1998}, (8) \cite{Bovino2011}, (9) \cite{Lepp2002}, (10) \cite{Mizusawa2005}, (11) \cite{McElroy2013}. 
\label{tab:continued}
\end{table}

The chemical network consists of 214 reactions, i.e., 107 forward and reverse pairs, among the following 23 species:
H, H$_2$, e$^-$, H$^+$, H$_2^+$, H$_3^+$, H$^-$, He, He$^+$, He$^{2+}$, HeH$^+$, 
D, HD, D$^+$, HD$^+$, D$^-$, Li, LiH, Li$^+$, Li$^-$, LiH$^+$, Li$^{2+}$, Li$^{3+}$.
The list of the reactions included and the references for the rate coefficients are shown in Table \ref{tab:chem_react}.
For a given forward reaction rate coefficient $k_{\rm fwd}$, that of the reverse reaction $k_{\rm rev}$ can be calculated 
from the principle of detailed balance:
\begin{equation}
k_{\rm rev} = k_{\rm fwd} K_{\rm eq}(T),
\label{eq:detailed_balance}
\end{equation}
where $K_{\rm eq}(T)$ is the equilibrium constant~\citep[e.g.,][]{Draine2011}.
For a reaction with $M$ reactants ${\rm R}_i$ and $N$ products ${\rm P}_i$, 
\begin{equation}
{\rm R}_1 + {\rm R}_2 + ... + {\rm R}_M \rightarrow {\rm P}_1 + {\rm P}_2 + ... + {\rm P}_N, 
\end{equation}
the equilibrium constant is given as 
\begin{equation}
\begin{split}
K_{\rm eq}(T) &= \left(\frac{2 \pi k_{\rm B} T}{h_{\rm P}^2}\right)^{\frac{3}{2}(M-N)} 
\left(\frac{m_{{\rm R}_1}...m_{{\rm R}_M}}{m_{{\rm P}_1}...m_{{\rm P}_N}}\right)^{\frac{3}{2}} \\
&\times \left(\frac{z({{\rm R}_1})...z({{\rm R}_M})}{z({{\rm P}_1})...z({{\rm P}_N})}\right) e^{-\Delta E/k_{\rm B} T},
\end{split}
\label{eq:eqb_const}
\end{equation}
where $z(i)$ is the partition function of the species $i$ and $\Delta E$ is the latent heat of the reaction, 
which is related with the ionization and dissociation energies of atoms and molecules $E(i)$ as 
$\Delta E \equiv (E({{\rm R}_1}) + ... + E({{\rm R}_M}))-(E({{\rm P}_1})+...+E({{\rm P}_N}))$.
Once the cloud becomes optically thick to the continuum, i.e., the continuum optical depth $\tau_{\rm cont} > 1$, 
the radiation field reaches the black body $B_\nu(T)$ inside and 
the rate coefficient of a photo-dissociation reaction $k_{\rm dissoc}$ is given by Eq. \eqref{eq:detailed_balance} with 
its reverse radiative association reaction coefficient $k_{\rm assoc}$. 
When the cloud is optically thin, i.e., $\tau_{\rm cont} < 1$, $k_{\rm dissoc}$ increases in proportion to the radiation intensity 
$J_\nu = (1-e^{-\tau_{\rm cont}})B_\nu(T)$.
We thus use the relation 
\begin{equation}
k_{\rm dissoc} = (1-e^{-\tau_{\rm cont}}) k_{\rm assoc} K_{\rm eq}(T).
\label{eq:detailed_balance_photo}
\end{equation}
For the partition function of H$_2$, we use the fitting formula by \cite{Popovas2016}.
For ${\rm H}_2^+, {\rm HeH}^+$, and ${\rm Li}$, those are taken from \cite{Barklem2016}, and 
for ${\rm HD}^+, {\rm H}_3^+, {\rm LiH}$, and ${\rm LiH}^+$ from the ExoMol database~\citep{Tennyson2012}\footnote{\url{http://exomol.com/data/molecules/}}, 
and for ${\rm HD}$ from the HITRAN database~\citep{Gamache2017}\footnote{\url{http://hitran.org/docs/iso-meta/}}.
For the other species, the values in the electronic ground state are take from the NIST-JANAF thermochemical table: $z({\rm H}) = 2, z(e) = 2, z({\rm H}^+) = 1, z({\rm H}^-) = 1, z({\rm He}) = 1, z({\rm He}^+) = 2, z({\rm He}^{2+}) = 1, z({\rm D}) = 2, z({\rm D}^+) = 1, z({\rm D}^-) = 1, z({\rm Li}^+) = 1, z({\rm Li}^-) = 1, z({\rm Li}^{2+}) = 2$, and $z({\rm Li}^{3+}) = 1$~\citep{Chase1998}\footnote{\url{https://janaf.nist.gov/}.}.
The ionization and dissociation energies of hydrogen- and helium-bearing species are taken from the KIDA database \footnote{\url{http://kida.obs.u-bordeaux1.fr/}}, 
those of deuterium-bearing species from the Active Thermochemical Table \footnote{\url{https://atct.anl.gov/Thermochemical Data/}}, 
and those of Li-bearing species from the NIST-JANAF Thermochemical Table, except for ${\rm LiH}^+$ from \cite{Stancil1996}.

\section{Thermal Evolution and Ionization Degree}
\label{sec:thermal_evolution}
We describe the result for the cases without external ionization in Section \ref{subsec:fiducial} and 
then see the external ionization effect in Section \ref{subsec:CR}. 
In Section \ref{subsec:minimal}, we present the minimal chemical network that correctly 
reproduces the ionization degree evolution during the pre-stellar collapse.  

\subsection{Fiducial cases with no external ionization source}
\label{subsec:fiducial}
\begin{figure}
\begin{center}
{\includegraphics[scale=1.1]{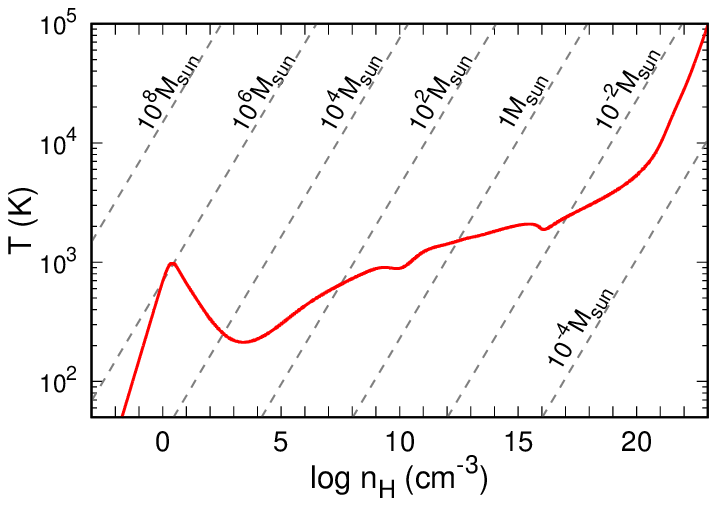}}\\
{\includegraphics[scale=1.1]{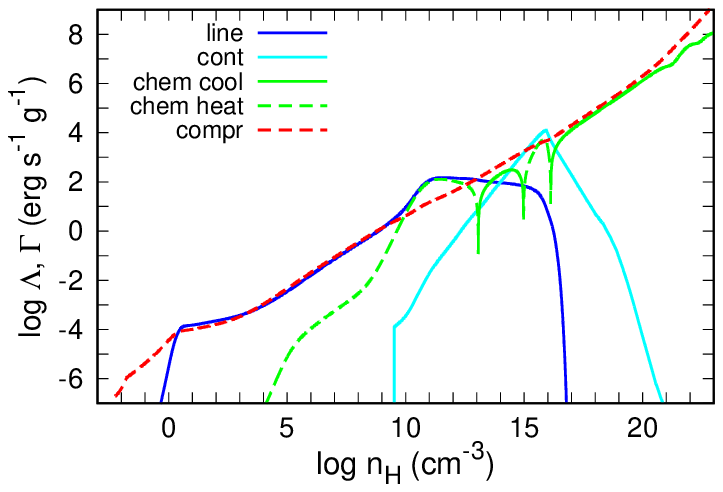}}
\caption{Temperature~(top) and cooling/heating rates~(bottom) as functions of the number density.
In the top panel, the grey dashed lines indicate the loci of constant Jeans mass.
In the bottom panel, blue, light blue, green solid lines correspond to the line, continuum, and chemical cooling, respectively,
and green and red dashed lines to the chemical and compressional heating, respectively.}
\label{fig:fig_nT}
\end{center}
\end{figure}

\begin{figure}
\begin{center}
\begin{tabular}{c}
{\includegraphics[scale=1.1]{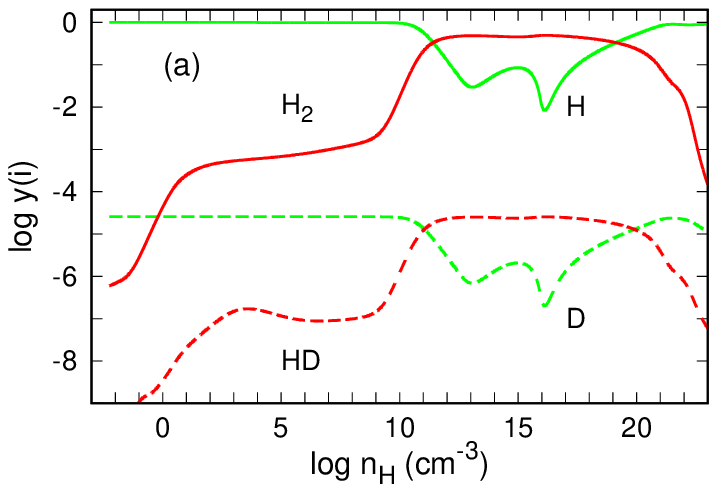}}\\
{\includegraphics[scale=1.1]{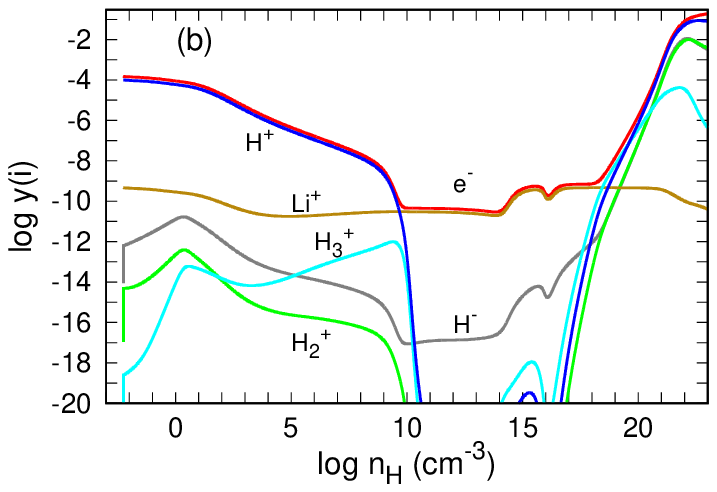}}\\
{\includegraphics[scale=1.1]{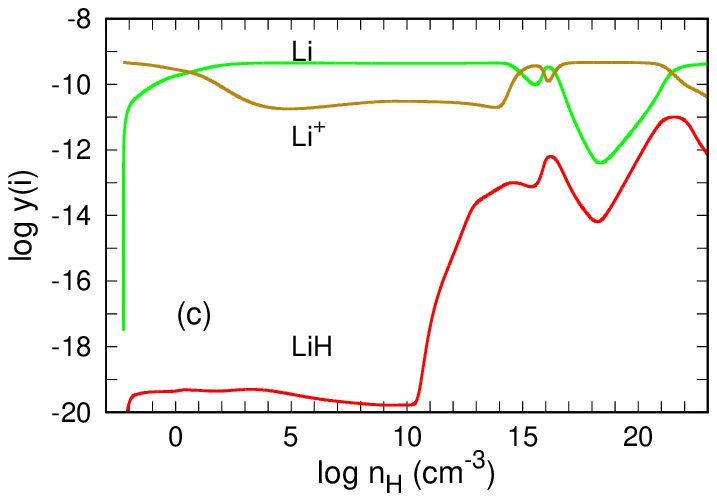}}
\end{tabular}
\caption{Fractional abundances of (a) H, H$_2$, D, and HD, (b) main charged species, and (c) Li, Li$^+$, and LiH as functions of the number density.}
\label{fig:fig_chem}
\end{center}
\end{figure}

\begin{figure}
\begin{center}
{\includegraphics[scale=1.1]{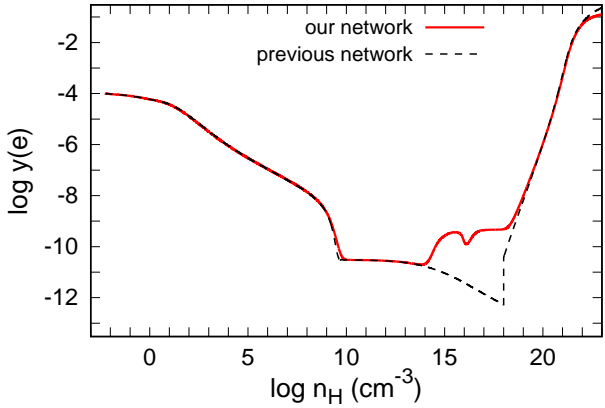}}
\caption{Comparison between the ionization degree calculated by using our chemical network~(red-solid line; same as the red line in Figure \ref{fig:fig_chem}b) and that calculated by using the previous network~(black-dashed line).}
\label{fig:fig_ele}
\end{center}
\end{figure}

We show the temperature and cooling/heating rates in a cloud as functions of the number density in the absence of external ionization sources in Figure \ref{fig:fig_nT}. 
Corresponding chemical abundances are shown in Figure \ref{fig:fig_chem} for 
(a) H, H$_2$, D, and HD, (b) main charged species, and (c) Li, Li$^+$, and LiH.

At the beginning ($\lesssim 1\ {\rm cm^{-3}}$), the fraction of H$_2$ is very low~($y({\rm H}_2) < 10^{-4}$), 
and the temperature increases adiabatically without efficient coolant~(Figures \ref{fig:fig_nT} and \ref{fig:fig_chem}a).
With the density increasing, H$_2$ formation proceeds via the H$^-$ channel~\citep{Peebles1968,Hirasawa1969}:
\begin{gather}
\label{eq:Hm_process_1}
{\rm H} + e \rightarrow {\rm H}^{-} + \gamma,\\
{\rm H}^{-} + {\rm H} \rightarrow {\rm H}_2 + e.
\label{eq:Hm_process_2}
\end{gather}
By the density of $\sim 10^4\ {\rm cm}^{-3}$, the H$_2$ fraction reaches $y({\rm H}_2) \sim 10^{-3}$, 
and the temperature decreases to $\sim 200\ {\rm K}$ via its cooling.
At $\gtrsim 10^4\ {\rm cm}^{-3}$, since the H$_2$ rotational levels reach the LTE and the cooling becomes inefficient, 
the temperature starts increasing again.  
At $\sim 10^{8}\ {\rm cm}^{-3}$, H$_2$ starts forming rapidly via the three-body reaction~\citep{Palla1983}:
\begin{equation}
{\rm H} + {\rm H} + {\rm H} \rightarrow {\rm H}_2 + {\rm H},
\label{eq:3b_H2}
\end{equation}
and eventually all the hydrogen becomes molecular by $\sim 10^{11}\ {\rm cm}^{-3}$.
Despite the elevated H$_2$ fraction, the temperature increase is interrupted only temporarily at $\sim 10^{10}\ {\rm cm}^{-3}$
since the enhanced cooling is compensated by the heating associated with the H$_2$ formation and 
also the H$_2$ lines become optically thick.
At $\sim 10^{14}-10^{16}\ {\rm cm}^{-3}$, the temperature slightly drops by the H$_2$ CIE cooling~\citep{Omukai1998}.
After the density exceeds $\sim 10^{16}\ {\rm cm}^{-3}$, the cloud becomes optically thick to the collision induced absorption, 
and the temperature increases monotonically with contraction.

Next, we see the fractional abundances of charged species as functions of the number density~(see Figure \ref{fig:fig_chem}b).
While the electron always remains the main negative charge carrier, 
the positive charge is successively carried by different cation species depending on the density: 
H$^+$ at $\lesssim 10^{10}\ {\rm cm}^{-3}$, Li$^+$ at $10^{10}-10^{18}\ {\rm cm}^{-3}$, 
H$_3^+$ at $10^{18}-10^{19.5}\ {\rm cm}^{-3}$, and H$^+$ at higher densities.
Until $\sim 10^{8}\ {\rm cm}^{-3}$, the ionization degree monotonically decreases as a result of 
the H$^+$ radiative recombination
\begin{equation}
{\rm H}^+ + e \rightarrow {\rm H} + \gamma.
\label{eq:hyd_recomb}
\end{equation}
At these densities, the ionization degree is set by the balance between the local free-fall time~($t_{\rm col}$) 
and the recombination time~($t_{\rm H, rec} = 1/k_{\rm H, rec} y(e) n_{\rm H}$): 
\begin{equation}
y(e) \simeq (k_{\rm H, rec} n_{\rm H} t_{\rm col})^{-1} \propto n_{\rm H}^{-1/2}.
\label{eq:ye_rec}
\end{equation}
At $\sim 10^{9}-10^{10}\ {\rm cm}^{-3}$, the increase of ${\rm H}_2$ fraction 
by the three-body reaction~(Eq. \ref{eq:3b_H2}) opens up another recombination channel via ${\rm H}_3^+$:  
\begin{equation}
\label{eq:H2p_form}
{\rm H} + {\rm H}^+ \rightarrow {\rm H}_2^+ + \gamma,
\end{equation}
\begin{equation}
{\rm H}_2^+ + {\rm H}_2 \rightarrow {\rm H}_3^+ + {\rm H},
\label{eq:H3p_form}
\end{equation}
and then ${\rm H}_3^+$ recombines by
\begin{equation}
{\rm H}_3^+ + e \rightarrow {\rm H}_2 + {\rm H}, \\
\end{equation}
or by
\begin{equation}
{\rm H}_3^+ + e \rightarrow 3{\rm H}.
\label{eq:H3p}
\end{equation}
As a result, the ionization degree drops exponentially at $\sim 10^{9}-10^{10}\ {\rm cm}^{-3}$. 
While ${\rm H}^+$ is the dominant cation species at $< 10^{10}\ {\rm cm^{-3}}$, 
the ${\rm Li}^+$ fraction is almost constant at $\sim 3 \times 10^{-11}$
since the Li$^+$ radiative recombination 
\begin{equation}
{\rm Li}^+ + e \rightarrow {\rm Li} + \gamma
\label{eq:lit_recomb}
\end{equation}
is balanced by the Li collisional ionization
\begin{equation}
{\rm Li} + {\rm H}^+ \rightarrow {\rm Li}^+ + {\rm H} +  \gamma.
\label{eq:lit_ionize}
\end{equation}
At $10^{10}-10^{14}\ {\rm cm}^{-3}$, since the Li$^+$ recombination time is much longer than the collapse time,
the ionization degree remains almost constant at $\sim 3 \times 10^{-11}$.
At higher densities $\gtrsim 10^{14}\ {\rm cm}^{-3}$, the lithium atom is photoionized 
\begin{equation}
{\rm Li} + \gamma \rightarrow {\rm Li}^{+} + e
\label{eq:lit_ionize}
\end{equation}
by trapped thermal photons due to the increasing temperature and continuum optical depth,
and the ionization degree is boosted by nearly two orders of magnitude.
The slight dip in the ionization degree at $\sim 10^{16}\ {\rm cm}^{-3}$ is caused by 
the temporal temperature drop at $\sim 10^{16}\ {\rm cm}^{-3}$ via the H$_2$ CIE cooling. 
At $> 10^{18}\ {\rm cm}^{-3}$, the temperature becomes high enough to ionize hydrogen and 
H$^+$ becomes the main cation thereafter.

Lithium is always mostly in the form of Li and Li$^+$, and LiH remains very minor with its fraction
at most $10^{-4}-10^{-3}$ of the total lithium fraction~(Figure \ref{fig:fig_chem}c). 
The main LiH formation channels are the following reactions:
\begin{equation}
{\rm Li} + {\rm H} \rightarrow {\rm LiH} + \gamma,
\end{equation}
and 
\begin{equation}
{\rm Li} + {\rm H}_2 \rightarrow {\rm LiH} + {\rm H},
\label{eq:LiH_form}
\end{equation}
at $\lesssim 10^{16}\ {\rm cm^{-3}}$.
The three-body association reactions proposed by \cite{Stancil1996}, 
\begin{equation}
\label{eq:3b_LiH_1}
{\rm Li} + {\rm H} + {\rm H} \rightarrow {\rm LiH} + {\rm H},
\end{equation}
and
\begin{equation}
{\rm Li} + {\rm H} + {\rm H}_2 \rightarrow {\rm LiH} + {\rm H}_2,
\label{eq:3b_LiH_2}
\end{equation}
only make minor contribution~\citep[see][]{Mizusawa2005}, 
and their highly uncertain rate coefficients do not affect the result. 
At higher densities, dissociation by the reaction
\begin{equation}
{\rm LiH} + {\rm H} \rightarrow {\rm Li} + {\rm H}_2, 
\label{eq:LiH_diss}
\end{equation}
balances the formation by the reactions above.  
Since the dissociation reaction is endothermic and fast, the equilibrium value of 
LiH is low and the lithium never becomes fully molecular.  

Finally, we see the difference in the ionization degree among our and previous models. 
We compare the ionization degree by our chemical network~(red-solid line; same as the red line in Figure \ref{fig:fig_chem}b) 
with that by the previous one~(black-dashed line) in Figure \ref{fig:fig_ele}. 
While we include the reverse reactions for all the 107 forward reactions, 
only 21 reverse reactions~(reactions H3, H5-7, He6, He9, He12, D1-3, D10, D17, D18, D20-22, D37-39, Li11, Li18 in Table \ref{tab:chem_react}) have been considered in the previous model by \cite{Maki2007}.
At $\sim 10^{14}-10^{18}\ {\rm cm}^{-3}$, the ionization degree is underestimated in the previous model 
by two to three orders of magnitude compared to our result. 
This is owing to their omission of the Li photoionization by thermal photons.  
Instead of Li photoionization at $\sim 10^{14}\ {\rm cm}^{-3}$,
the ionization degree continues decreasing due to the Li$^+$ recombination in the previous model.
In addition, the ionization degree jumps discontinuously at $10^{18}\ {\rm cm}^{-3}$
when the chemical abundance is switched to the equilibrium value by hand. 
The ionization degree in our model, on the other hand, smoothly reaches the equilibrium value.  

\subsection{Effect of external ionization}
\label{subsec:CR}

\begin{figure}
\begin{center}
\begin{tabular}{c}
{\includegraphics[scale=1.1]{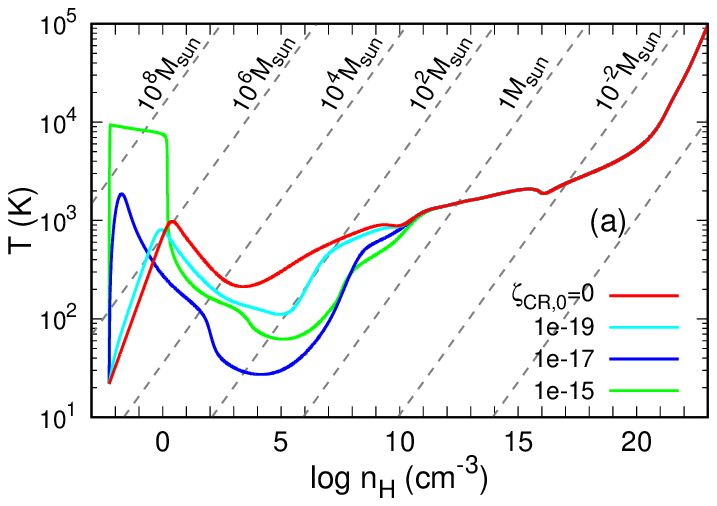}} \\
{\includegraphics[scale=1.1]{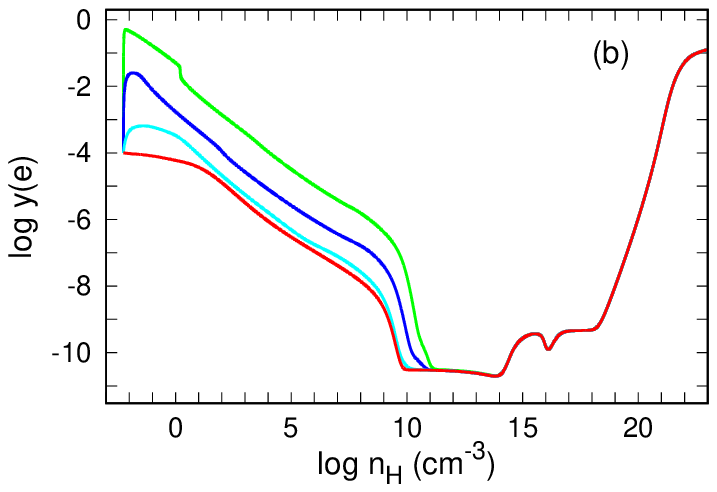}} \\
{\includegraphics[scale=1.1]{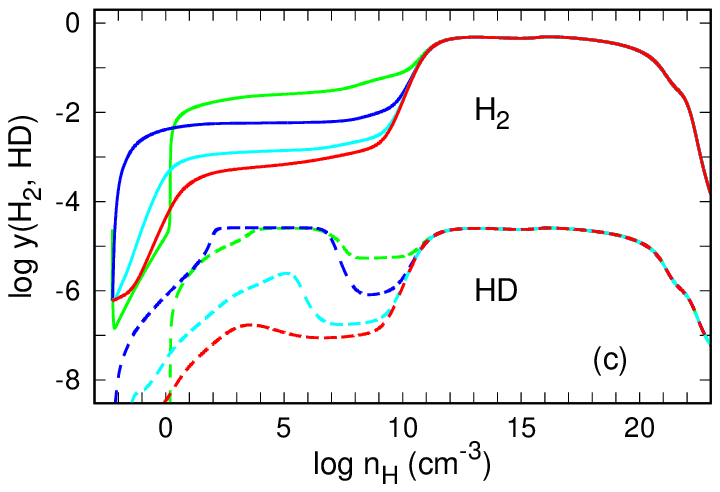}}
\end{tabular}
\caption{CR effects on (a) thermal evolution and fractional abundances of (b) electron, (c) H$_2$, and HD.
Colors show the results with $\zeta_{\rm CR, 0}=0$~(red), $10^{-19}$~(light blue), $10^{-17}$~(blue), and $10^{-15}\ {\rm s}^{-1}$~(green), respectively. Note that the results with $\zeta_{\rm CR, 0}=0$~(the red color) are the same as those in Figures \ref{fig:fig_nT}a, \ref{fig:fig_chem}a, and \ref{fig:fig_ele}.}
\label{fig:fig_cr_nT}
\end{center}
\end{figure}

So far, we have not considered the effect of the external ionization and heating by 
cosmic-ray~(CR) or X-ray sources.
Already in the epoch of first galaxy formation, however, CR particles could have been 
generated in Pop III supernova remnants~(SNRs), and X-rays might be emitted from early populations of 
high-mass X-ray binaries or micro-quasars~\citep[e.g.,][]{Glover2003,Stacy2007}.
Here, we consider CR irradiation as a representative of external ionization/heating 
and study its effect on the ionization degree in the primordial gas clouds. 

The CR ionization rate $\zeta_{\rm CR, 0}$ in the high-$z$ universe is still largely unknown~\citep[e.g.,][]{Stacy2007, Nakauchi2014}.
Since the cloud under consideration is in the primordial composition, 
star formation has not proceeded remarkably and no strong CR source would be present in the same halo.  
On the other hand, star formation and the CR injection can be vigorous in starburst galaxies, and 
the CR incidation rate onto our cloud can be high if such galaxies locate nearby.
Considering this large uncertainty, here we study the cases with a wide range of the CR ionization rate 
$\zeta_{\rm CR, 0} = 0, 10^{-19}, 10^{-17}, 10^{-15}\ {\rm s}^{-1}$, encompassing 
the recent estimation in the Galactic disk of $2.3 \times 10^{-16}\ {\rm s}^{-1}$~\citep{Neufeld2017}.
Note that the renewed Galactic value is more than an order of magnitude higher from the previous conventional 
value $\sim 10^{-17}\ {\rm s}^{-1}$~\citep{Spitzer1968}.

The incident CR particles lose their energy by ionizing the neutral atoms and molecules during propagation in the 
cloud, and cannot reach the very dense central part. 
We include the shielding effect for the CR ionization rate inside the cloud $\zeta_{\rm CR}$ as~\citep{Nakano1986}:
\begin{equation}
\zeta_{\rm CR} = \zeta_{\rm CR, 0} \exp\left(-\frac{N_{\rm H}}{4.3 \times 10^{25}\ {\rm cm}^{-2}}\right), 
\label{eq:CR_intensity}
\end{equation}
where $N_{\rm H}$ is the column density of the cloud.
We calculate the heating rate by CR ionization~($\Gamma_{\rm CR}$) assuming that 3.4 eV of heat is given to the gas 
per ionization~\citep{Spitzer1969}.
The list of the CR ionization/dissociation reactions, and the CR-induced photo-reactions is also presented 
in Table \ref{tab:chem_react}.

The temperature and the fractional abundances of electron, H$_2$, and HD in the cases with the CR incidation are 
shown as functions of the number density in Figure \ref{fig:fig_cr_nT} (a-c).
The different colors correspond to the cases with the different CR ionization rate: 
$\zeta_{\rm CR, 0}=0$~(red), $10^{-19}$~(light blue), $10^{-17}$~(blue), 
and $10^{-15}\ {\rm s}^{-1}$~(green), respectively.
The CRs affect the thermal evolution in two notable ways.
At very low densities $\lesssim 1\ {\rm cm}^{-3}$, due to the CR ionization heating, the temperature becomes higher than
the case without CRs~(Figure \ref{fig:fig_cr_nT}a).
On the other hand, at higher densities $\gtrsim 1\ {\rm cm}^{-3}$, 
the CR ionization enhances the H$_2$ abundance and its cooling, thereby lowering the temperature 
below that without CRs.
Those CR effects are observed only below the density $\sim 10^{11}\ {\rm cm}^{-3}$ 
as the CRs are completely shielded at higher densities.

When the CR intensity is high enough $\zeta_{\rm CR, 0} \geq 10^{-19}\ {\rm s}^{-1}$, 
the ionization degree $y(e)$ is determined by the balance between the hydrogen recombination and the CR ionization: 
\begin{equation}
y(e) \sim \left(\frac{\zeta_{\rm CR, 0}}{k_{\rm H, rec} n_{\rm H}}\right)^{1/2},
\label{eq:ye_cr}
\end{equation}
and increases with the CR intensity~(Figure \ref{fig:fig_cr_nT}b).
Comparing this with the ionization degree without CRs~(Eq. \ref{eq:ye_rec}), 
we can see that the CR ionization controls the ionization rate for $\zeta_{\rm CR, 0} \gtrsim 10^{-19}\ {\rm s}^{-1}$, 
consistent with our result. 
This enhanced ionization fraction promotes the H$_2$ formation via the H$^-$ channel, which is catalyzed by 
electrons~(Eqs. \ref{eq:Hm_process_1} and \ref{eq:Hm_process_2}) and the cloud cools to lower temperature. 
Once the temperature falls below $\sim150$ K, another coolant, HD, kicks in 
since the HD formation reactions~(Figure \ref{fig:fig_cr_nT}c dashed lines)
\begin{gather}
{\rm D} + {\rm H}^{+} \rightarrow {\rm D}^{+} + {\rm H}, \\
{\rm D}^+ + {\rm H}_2 \rightarrow {\rm H}^+ + {\rm HD},
\label{eq:HD_form}
\end{gather}
are exothermic and HD formation proceeds in cold environments.
Thanks to the HD cooling, the temperature 
drops even more to a few 10 K~\citep[e.g.,][]{Omukai2005,Nagakura2005}.
At $\sim 10^{5}-10^{6}\ {\rm cm}^{-3}$, the HD rotational levels reach the LTE and the temperature starts increasing again.
This increase is steeper than the case without CRs because of the HD dissociation accompanying the temperature increase and loss of its cooling. 
All the thermal tracks converge to that without CRs ($\zeta_{\rm CR, 0}=0$) at $\sim 10^{8}-10^{10}\ {\rm cm}^{-3}$,
where the gas is heated by the H$_2$ formation via the three-body reactions. 

\subsection{Minimal chemical network}
\label{subsec:minimal}
\begin{figure}
\begin{center}
\begin{tabular}{c}
{\includegraphics[scale=1.1]{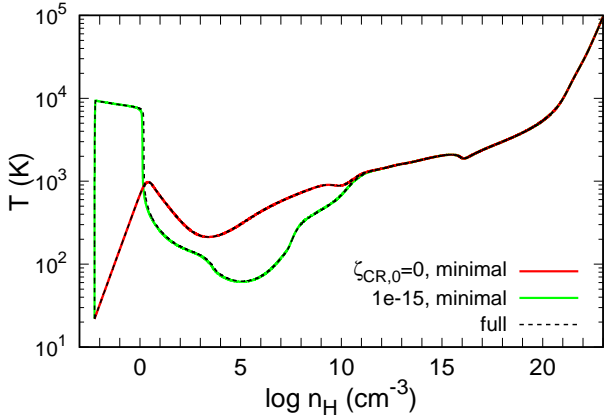}}
\end{tabular}
\caption{Thermal evolution calculated with the minimal~(solid lines) and full~(black-dashed lines) chemical networks. Colors indicate the models with $\zeta_{\rm CR, 0} = 0$~(red), and $10^{-15}\ {\rm s}^{-1}$~(green), respectively.}
\label{fig:nT_red}
\end{center}
\end{figure}

\begin{figure}
\begin{center}
\begin{tabular}{c}
{\includegraphics[scale=1.0]{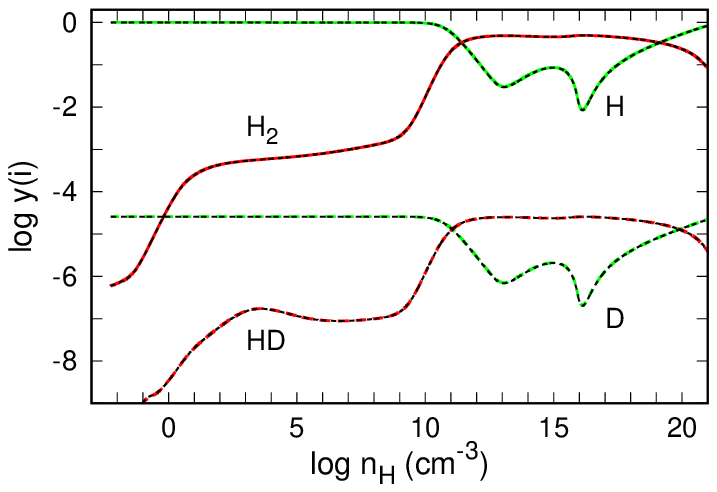}}\\
{\includegraphics[scale=1.0]{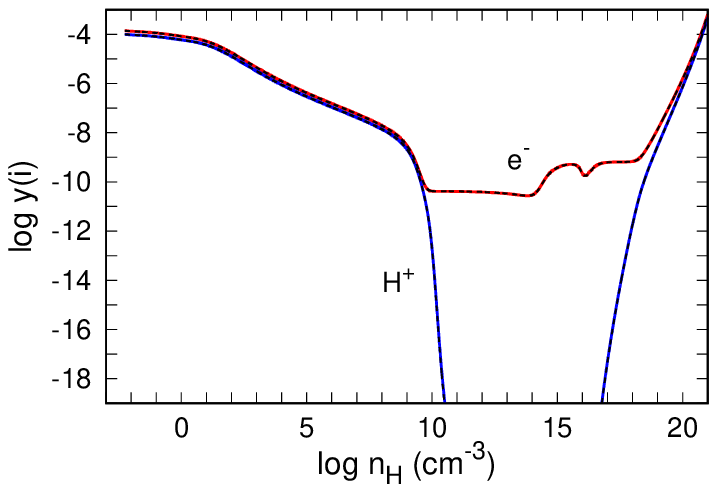}}\\
{\includegraphics[scale=1.0]{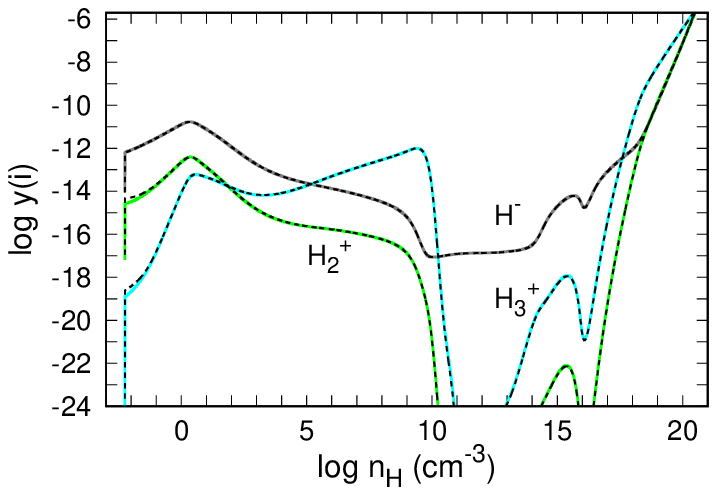}}\\
{\includegraphics[scale=1.0]{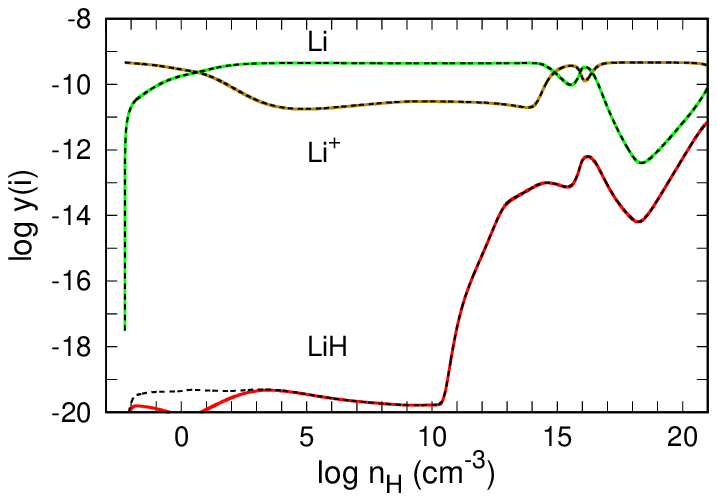}}
\end{tabular}
\caption{Fractional abundances of species calculated with the minimal~(solid lines) and full~(black dashed lines) chemical networks.
The fiducial case with no external ionization, $\zeta_{\rm CR, 0} = 0$.}
\label{fig:chem_red_CR00}
\end{center}
\end{figure}

\begin{figure}
\begin{center}
\begin{tabular}{c}
{\includegraphics[scale=1.0]{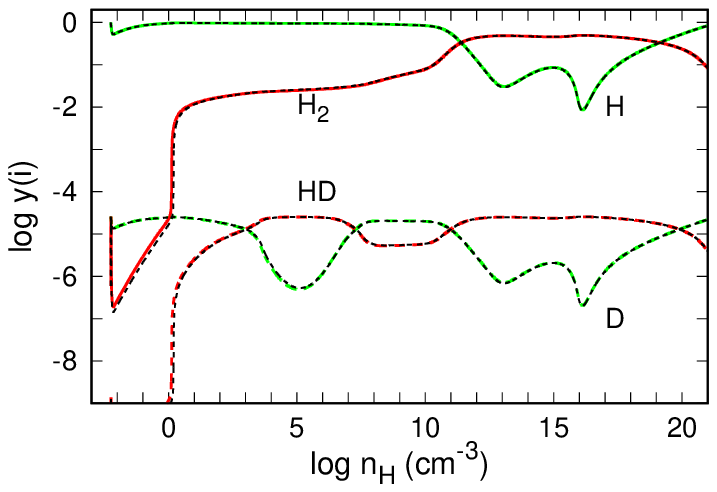}}\\
{\includegraphics[scale=1.0]{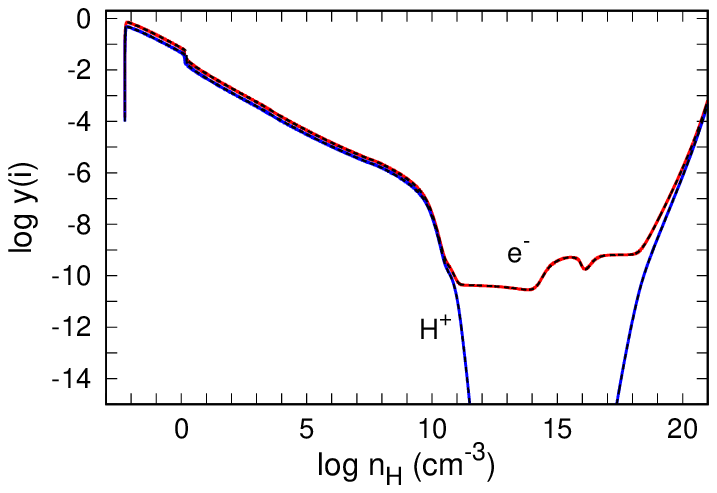}}\\
{\includegraphics[scale=1.0]{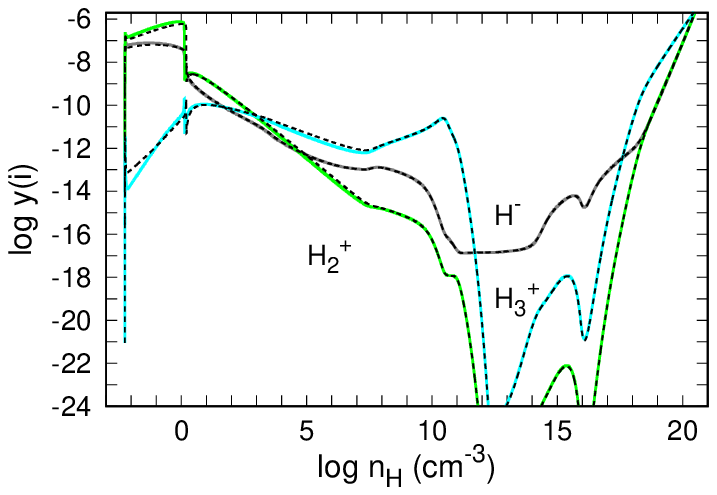}}\\
{\includegraphics[scale=1.0]{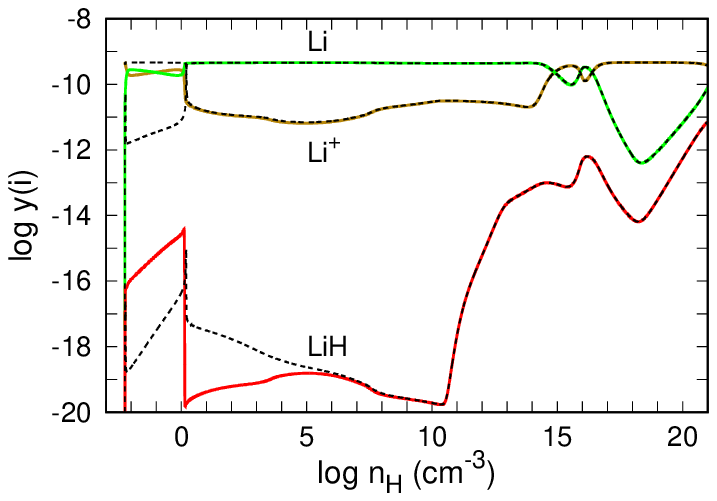}}
\end{tabular}
\caption{The same as Figure \ref{fig:chem_red_CR00}, but for the external ionization rate of $\zeta_{\rm CR, 0} = 10^{-15}\ {\rm s}^{-1}$.}
\label{fig:chem_red_CR12}
\end{center}
\end{figure}

\begin{table}
\caption{Minimal chemical reactions.}
\begin{center}
{\begin{tabular}{cl}
\hline
Number & Reaction \\
\hline
H1 & $ {\rm H}^+     +   e      \rightleftharpoons     {\rm H}        +    \gamma $       \\
H2 & $ {\rm H}         +   e      \rightleftharpoons     {\rm H}^{-}  +     \gamma $      \\
H3 & $ {\rm H}^{-}   + {\rm H}\rightleftharpoons {\rm H}_2 + e $    \\
H4 & $ {\rm H} + {\rm H}^{+}\rightleftharpoons {\rm H}_2^{+} + \gamma $\\
H5 & $ {\rm H}_2^{+} + {\rm H}\rightleftharpoons {\rm H}_2 + {\rm H}^{+} $ \\
H6 & $ 3{\rm H}\rightleftharpoons {\rm H}_2 + {\rm H} $ \\
H7 & $ 2{\rm H}_2\rightleftharpoons 2{\rm H} + {\rm H}_2 $\\
H8 & $ {\rm H}_2^+ + {\rm H}_2 \rightleftharpoons {\rm H}_3^+ + {\rm H} $ \\ 
H9 & $ {\rm H}_3^+ + e \rightleftharpoons {\rm H}_2 + {\rm H} $ \\
H10 & $ {\rm H}_3^+ + e \rightleftharpoons 3{\rm H}$\\
D1 & $ {\rm D} + {\rm H}^{+} \rightleftharpoons {\rm D}^{+} + {\rm H} $\\
D2 & $ {\rm D} + {\rm H}_2 \rightleftharpoons {\rm H} + {\rm HD} $\\ 
D3 & $ {\rm D}^+ + {\rm H}_2 \rightleftharpoons {\rm H}^+ + {\rm HD} $\\ 
Li1 & $ {\rm Li}^{+} + e \rightleftharpoons {\rm Li} + \gamma $ \\
Li2 & $ {\rm Li}^{+} + {\rm H}^{-} \rightleftharpoons {\rm Li} + {\rm H}$ \\
Li3 & $ {\rm Li} + {\rm H}^{+} \rightleftharpoons {\rm Li}^+ + {\rm H} + \gamma$ \\
Li4 & $ {\rm LiH} + {\rm H} \rightleftharpoons {\rm Li} + {\rm H}_2$ \\
Li5 & $ {\rm Li} + {\rm H} \rightleftharpoons {\rm LiH} + \gamma$ \\
H11$^\ast$ & $ {\rm H}^- + {\rm H}^+ \rightleftharpoons 2{\rm H}$ \\
H12$^\ast$ & $ {\rm H}_2^+ + e \rightleftharpoons 2{\rm H}$ \\
H13$^\ast$ & $ {\rm H}^- + {\rm H} \rightleftharpoons 2{\rm H}$ + e\\
He1$^\ast$ & $ {\rm He}^+     +   e      \rightleftharpoons     {\rm He}        +    \gamma $       \\
He2$^\ast$ & $ {\rm He}^+ + {\rm H}_2 \rightleftharpoons {\rm H}^+ + {\rm H} + {\rm He} $ \\ 
\hline
\end{tabular}}
\end{center}
{\bf Notes.} Reactions with asterisks are needed only in the presence of CRs.
\label{tab:chem_react_red}
\end{table}

The full chemical network in this paper involves a number of species (23) and reactions (214).
Although the ionization degree can be followed correctly, it is computationally too costly 
to incorporate the full network into, e.g., the multi-dimensional numerical simulations. 
With such future applications in mind, we here construct the minimal chemical network that can correctly reproduce 
the evolution of the temperature and ionization degree~(Table \ref{tab:chem_react_red}).

We find that the evolution can be followed correctly only with 36 reactions~(18 forward and reverse pairs of 1st-18th rows in Table \ref{tab:chem_react_red}) among 13 species~(H, H$_2$, e$^-$, H$^+$, H$_2^+$, H$_3^+$, H$^-$, D, HD, D$^+$, Li, LiH, Li$^+$; Figure \ref{fig:nT_red} red line and Figure \ref{fig:chem_red_CR00}) in the case without CRs.
The temperature tracks and chemical abundances calculated by the minimal~(colored-solid lines) and by 
the full~(black-dashed lines) networks are shown together in Figures \ref{fig:nT_red} and \ref{fig:chem_red_CR00} for comparison.
We can see they agree very well except the LiH fraction at $\lesssim 10^{3}\ {\rm cm}^{-3}$, 
where LiH is very minor species anyway and is not relevant to the thermal evolution and 
the abundances of the other species.

When the CR is present, we need to add two more species (He and He$^+$) and 10 more reactions 
(19th-23rd rows in Table \ref{tab:chem_react_red}) to the above network. 
The comparison with the full network is presented in Figures \ref{fig:nT_red} and \ref{fig:chem_red_CR12} for the case with 
$\zeta_{\rm CR, 0} = 10^{-15}\ {\rm s}^{-1}$.
The additional reactions are needed to reproduce the evolution at $\lesssim 10\ {\rm cm}^{-3}$, 
where the temperature is initially elevated to $\sim$ 10000 K via the CR heating and 
subsequently drops via the H$_2$ cooling~(Figure \ref{fig:nT_red} green line and Figure \ref{fig:chem_red_CR12}).
Although there are some deviations in the Li chemistry from the full network, i.e., 
Li and Li$^+$ at $\lesssim 1\ {\rm cm}^{-3}$, 
and LiH at $\lesssim 10^{6}\ {\rm cm}^{-3}$~(Figure \ref{fig:chem_red_CR12} bottom panel),
they do not affect the thermal and ionization-degree evolution at all.

\section{Magnetic field dissipation in the primordial gas}
\label{sec:magnetic}
\subsection{Magnetic diffusivity}
\label{subsec:resistivity}
Owing to the low ionization degree in star-forming clouds, 
the magnetic field may decouple from the gas via the so-called non-ideal MHD effects. 
The field evolution in such a case is described by the following induction equation~\citep[e.g.,][]{Wardle2007,Tsukamoto2016}:
\begin{equation}
\begin{split}
\frac{\partial \bm{B}}{\partial t} &= \nabla \times (\bm{{\rm v}}_{\rm n} \times \bm{B}) \\
& - \nabla \times \left[\eta_{\rm Ohm} \left(\nabla \times \bm{B} \right) + \eta_{\rm Hall} \left(\nabla \times \bm{B} \right) \times \bm{e}_B \right.\\
&\left. \quad \quad \quad - \eta_{\rm ambi} \left((\nabla \times \bm{B}) \times \bm{e}_B \right) \times \bm{e}_B \right],
\end{split}
\label{eq:diff_eq}
\end{equation}
where $\bm{{\rm v}}_{\rm n}$ is the velocity of neutral components and $\bm{e}_B \equiv \bm{B}/|\bm{B}|$ the unit vector along the field line.
In the right hand side~(RHS), the first term corresponds to the flux conservation, and the second, third, and fourth to the magnetic dissipation due to the Ohmic loss, Hall effect, and ambipolar diffusion, respectively.
Individual diffusivity coefficients, $\eta_{\rm ambi}, \eta_{\rm Ohm}$, and $\eta_{\rm Hall}$, are given by~\citep[e.g.,][]{Wardle1999}
\begin{equation}
\eta_{\rm ambi} = \frac{c^2}{4 \pi} \frac{\sigma_{\rm P}}{\sigma_{\rm P}^2+ \sigma_{\rm H}^2} - \eta_{\rm Ohm},
\label{eq:eta_ambi}
\end{equation}
\begin{equation}
\eta_{\rm Ohm} = \frac{c^2}{4 \pi \sigma_{\rm O}},
\label{eq:eta_ohm}
\end{equation}
\begin{equation}
\eta_{\rm Hall} = \frac{c^2}{4 \pi} \frac{\sigma_{\rm H}}{\sigma_{\rm P}^2+ \sigma_{\rm H}^2},
\label{eq:eta_hall}
\end{equation}
where the Pedersen, Hall, and Ohmic conductivities, $\sigma_{\rm P}, \sigma_{\rm H}$, and $\sigma_{\rm O}$, are
\begin{equation}
\sigma_{\rm P} = \left(\frac{c}{B}\right)^2 \sum_\nu \frac{\rho_\nu \tau_\nu \omega_\nu^2}{1 + \tau_\nu^2 \omega_\nu^2},
\label{eq:sig_ped}
\end{equation}
\begin{equation}
\sigma_{\rm H} = \left(\frac{c}{B}\right)^2 \sum_\nu \frac{q_\nu}{|q_\nu|} \frac{ \rho_\nu \omega_\nu}{1 + \tau_\nu^2 \omega_\nu^2},
\label{eq:sig_hall}
\end{equation}
\begin{equation}
\sigma_{\rm O} = \left(\frac{c}{B}\right)^2 \sum_\nu \rho_\nu \tau_\nu \omega_\nu^2.
\label{eq:sig_ohm}
\end{equation}
Here the subscript `$\nu$' stands for the charged species whose charge $q_\nu$, mass density $\rho_\nu = m_\nu n_\nu$, 
and cyclotron frequency $\omega_\nu = e |q_\nu| B / m_\nu c$.
The collision time between the charged and neutral particles $\tau_\nu$ is given by~\citep{Nakano1986}
\begin{equation}
\tau_\nu^{-1} = \sum_{\rm n} \tau_{\nu, {\rm n}}^{-1} = \sum_{\rm n} \frac{\mu_{\nu, {\rm n}} n_{\nu} n_{\rm n} <\sigma {\rm v}>_{\nu, {\rm n}}}{\rho_\nu},
\label{eq:tau_nu}
\end{equation}
where the subscript `n' stands for H, H$_2$, and He, $\mu_{\nu, {\rm n}}$ the reduced mass, and $<\sigma {\rm v}>_{\nu, {\rm n}}$ the collisional rate coefficient.
For collisions between e-H, e-H$_2$, e-He, H$^+$-H, H$^+$-H$_2$, H$^+$-He, and H$_3^+$-H$_2$, $<\sigma {\rm v}>_{\nu, {\rm n}}$ are taken from \cite{Pinto2008}, 
and for the others from \cite{Osterbrock1961}.

\begin{figure}
\begin{center}
\begin{tabular}{c}
{\includegraphics[scale=1.1]{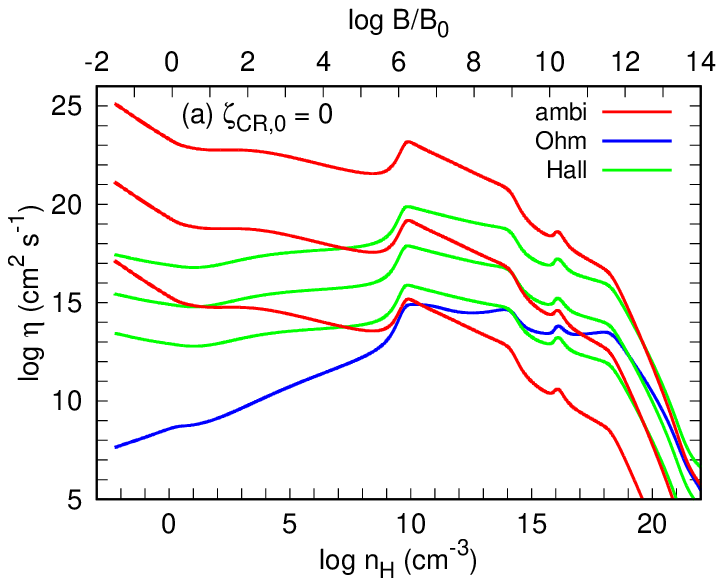}} \\
{\includegraphics[scale=1.1]{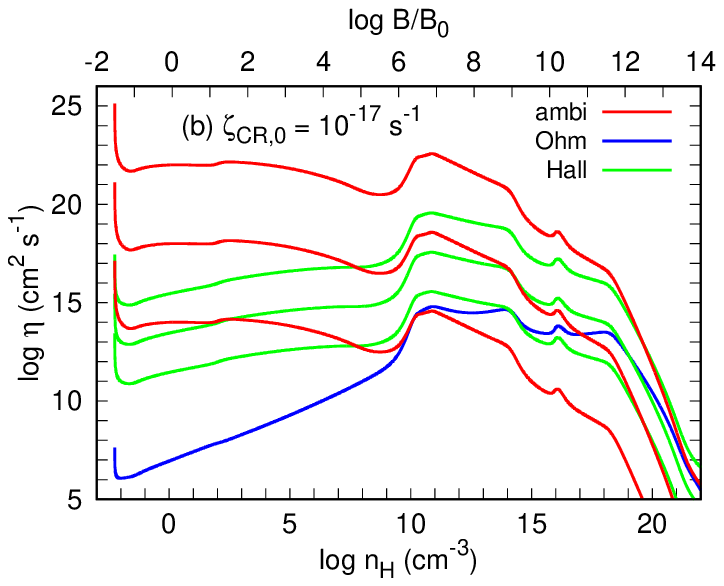}} \\
\end{tabular}
\caption{Magnetic diffusivities, $\eta_{\rm ambi}$~(red), $\eta_{\rm Ohm}$~(blue), and $\eta_{\rm Hall}$~(green) as functions of the number density for the CR ionization rates of $\zeta_{\rm CR, 0} = 0$~(top) and $10^{-17}\ {\rm s}^{-1}$~(bottom, respectively).
The magnetic field is assumed to be amplified as $B(n_{\rm H}) = B_0 n_{\rm H}^{2/3}$ with $B_0=(1, 10^{-2}, 10^{-4})\ \mu{\rm G}$ from top to bottom lines.
Note that the Ohmic diffusivity $\eta_{\rm Ohm}$~(blue) does not depend on the field strength and is shown by a single line.}
\label{fig:res}
\end{center}
\end{figure}

First, we see the dependence of the magnetic diffusivities $\eta_{\rm ambi}$, $\eta_{\rm Ohm}$, and $\eta_{\rm Hall}$ 
on the magnetic field strength and ionization degree, which is shown as a function of the number density
in Figure \ref{fig:res} for the cases with $\zeta_{\rm CR,0}=0$~(top) and $10^{-17}\ {\rm s}^{-1}$~(bottom panels).
We consider three different initial field strengths $B_0=(1, 10^{-2}, 10^{-4})\ \mu{\rm G}$ at $n_{\rm H} = 1\ {\rm cm}^{-3}$ and 
assume the field to increase with density as $B(n_{\rm H}) = B_0 n_{\rm H}^{2/3}$, which is the case in the spherical contraction without the magnetic flux loss. 
Since the diffusivities can be written as 
\begin{equation}
\eta_{\rm ambi} \simeq \frac{1}{4 \pi \mu_{\rm i, n}} \left(\frac{B}{n_{\rm H}}\right)^2 \left(<\sigma {\rm v}>_{{\rm i}, {\rm n}} y({\rm n}) y(e)\right)^{-1}, 
\label{eq:eta_ambi_app}
\end{equation}
\begin{equation}
\eta_{\rm Ohm} \simeq \frac{m_e c^2}{4 \pi e^2} <\sigma {\rm v}>_{e, {\rm n}} y({\rm n}) y(e)^{-1},
\label{eq:eta_ohm_app}
\end{equation}
\begin{equation}
\eta_{\rm Hall} \simeq \frac{c}{4 \pi e} \frac{B}{n_{\rm H}} y(e)^{-1},
\label{eq:eta_hall_app}
\end{equation}
by considering only the contributions from the major ionic and neutral species~(i and n) in Eqs. (\ref{eq:sig_ped}-\ref{eq:tau_nu}), 
$\eta_{\rm ambi}$ (red lines) and $\eta_{\rm Hall}$ (green) vary with $B$ as $\propto B^2$ and $\propto B$, respectively, at a given density, 
while $\eta_{\rm Ohm}$~(blue) does not depend on the field strength and is represented by a single line.
Note also that the diffusivities are inversely proportional to the ionization degree.
The enhanced electron fraction by CR ionization at $\lesssim 10^{11}\ {\rm cm}^{-3}$ 
makes the magnetic diffusivities smaller in case of $\zeta_{\rm CR,0}=10^{-17}\ {\rm s}^{-1}$ (Fig. \ref{fig:res} bottom panel).

Next, we examine the roles of the three non-ideal MHD effects in the field dissipation.
With Amp\`ere's law $\nabla \times \bm{B} = (4 \pi/c) \bm{J}$, Eq. \eqref{eq:diff_eq} can be rewritten as
\begin{equation}
\begin{split}
\frac{\partial \bm{B}}{\partial t} &= \nabla \times \left((\bm{{\rm v}}_{\rm n}+\bm{{\rm v}}_{\rm Hall}) \times \bm{B}\right) \\
& - \nabla \times \left[\left(\eta_{\rm ambi} + \eta_{\rm Ohm}\right) \left(\nabla \times \bm{B} \right)\right] \\
& + \nabla \times \left[\eta_{\rm ambi} \left(\frac{4 \pi}{c} \bm{J} \cdot \bm{e}_B \right) \bm{e}_B\right],
\end{split}
\label{eq:diff_eq2}
\end{equation}
where
\begin{equation}
\bm{{\rm v}}_{\rm Hall} = - \frac{4 \pi}{c} \eta_{\rm Hall} \frac{\bm{J}}{|\bm{B}|}.
\label{eq:v_hall}
\end{equation}
We assume that the cloud is threaded by a poloidal field~($\bm{B} = B \bm{e}_{B}$) produced by the toroidal current 
in the equatorial plane~($\bm{J} = J \bm{e}_{\phi}$), which leads to $\bm{J} \cdot \bm{B} = 0$ and the vanishing of the last term in the RHS of Eq. \eqref{eq:diff_eq2}.
Note also that while the Hall term~$\nabla \times (\bm{{\rm v}}_{\rm Hall} \times \bm{B}$) causes the field lines to drift 
from the neutral gas in the azimuthal direction with velocity $\bm{{\rm v}}_{\rm Hall}$~(Eq. \ref{eq:v_hall}), 
it does not remove the magnetic flux from the cloud~\citep[e.g.,][]{Tsukamoto2017}.
Therefore, the magnetic flux is lost only by the second term in the RHS of Eq. \eqref{eq:diff_eq2}, i.e., ambipolar diffusion and Ohmic loss.

\begin{figure}
\begin{center}
{\includegraphics[scale=1.1]{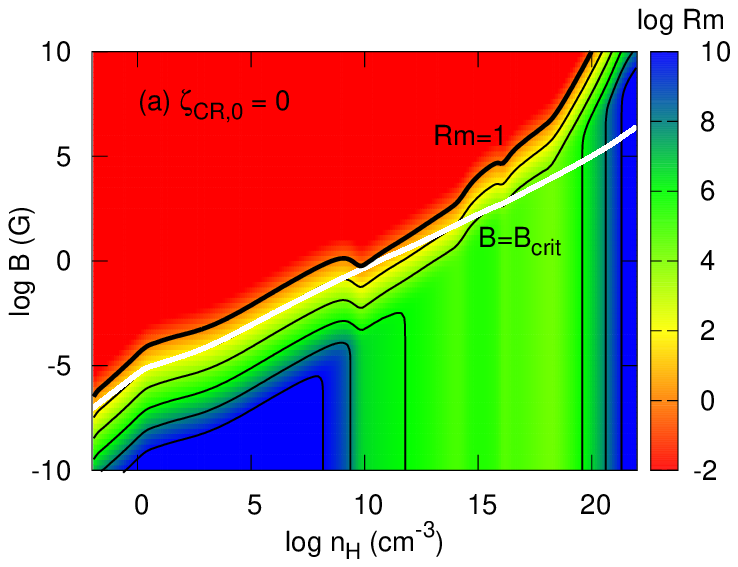}} \\
{\includegraphics[scale=1.1]{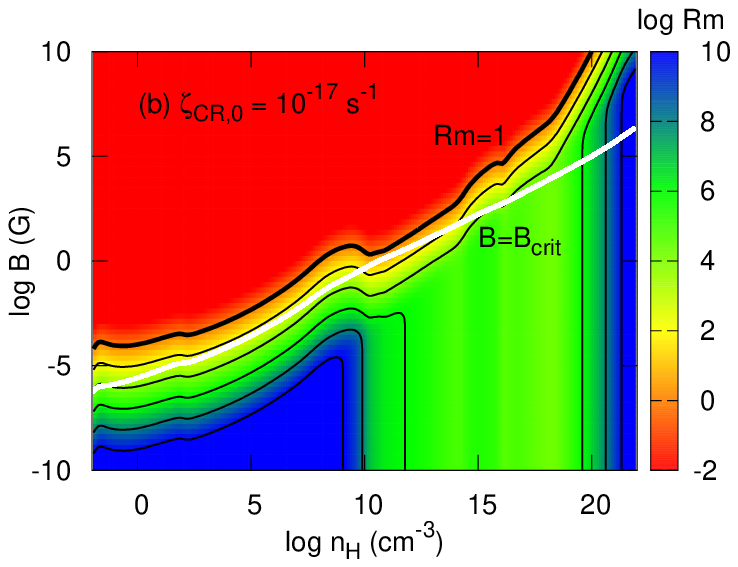}} 
\caption{Contour map of the magnetic Reynolds number ${\rm Rm}(\lambda_{\rm J})$ for the CR ionization rates of $\zeta_{\rm CR, 0} = 0$~(top) and $10^{-17}\ {\rm s}^{-1}$~(bottom, respectively).
Contours are plotted from $\log {\rm Rm}(\lambda_{\rm J}) = 0$ to 10 with the interval of 2~(black lines).
Magnetic field dissipates above the thick-black lines of ${\rm Rm}(\lambda_{\rm J})=1$.
The critical magnetic field strength $B_{\rm crit}$ is indicated by the white lines.}
\label{fig:Rm_rev}
\end{center}
\end{figure}
\begin{figure*}
\begin{center}
{\includegraphics[scale=1.1]{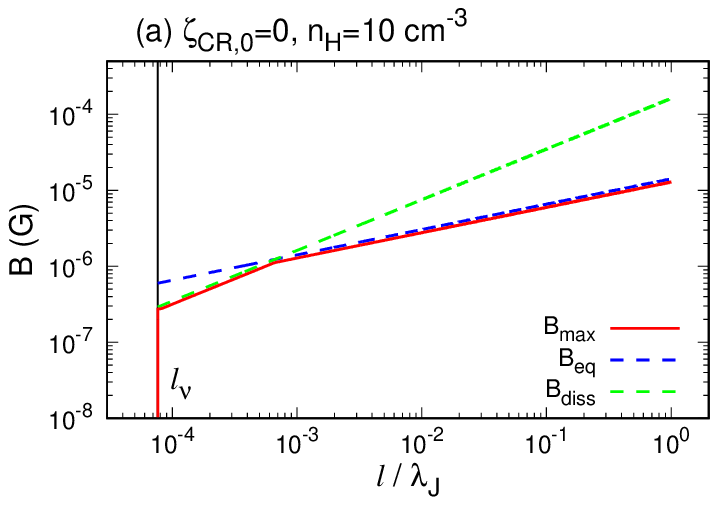}} 
{\includegraphics[scale=1.1]{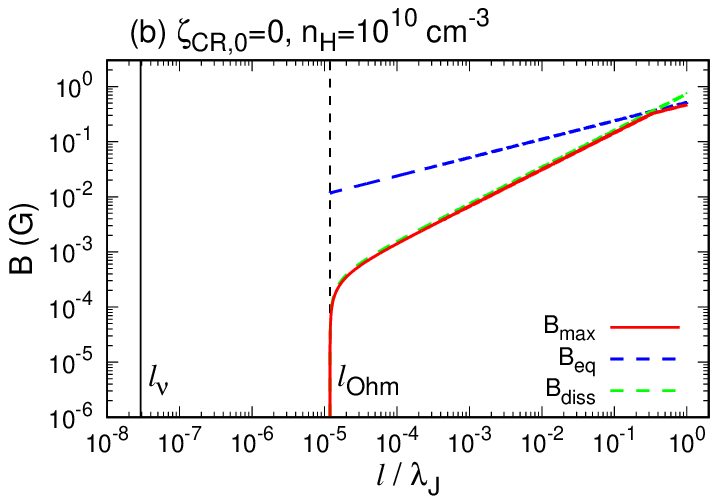}} \\
{\includegraphics[scale=1.1]{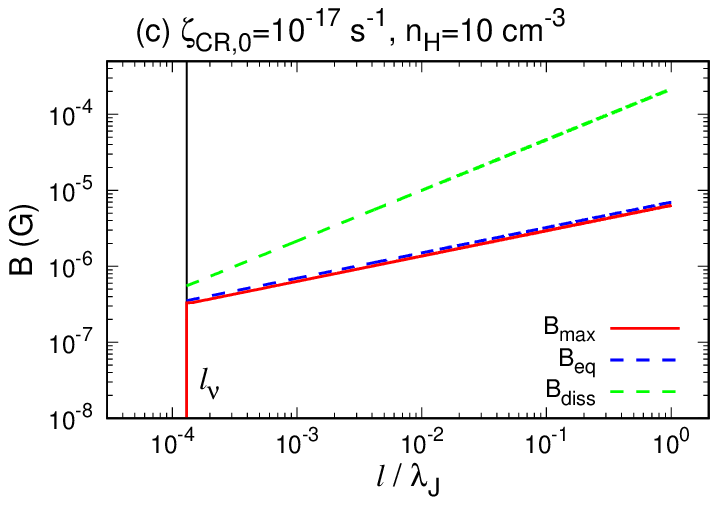}} 
{\includegraphics[scale=1.1]{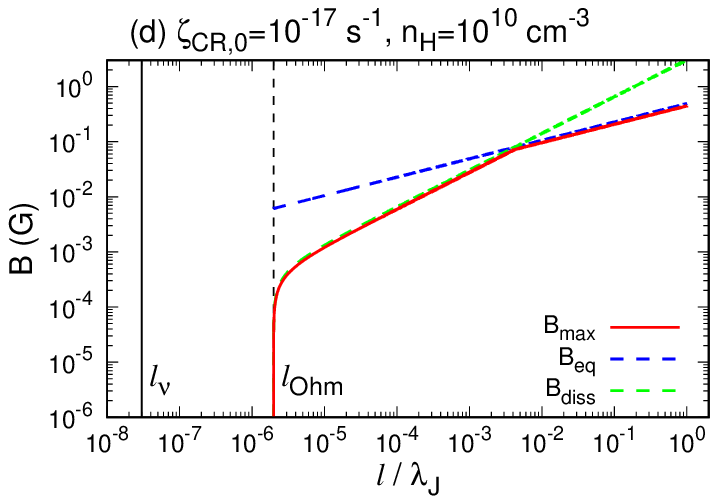}} \\
\caption{Maximum magnetic field strength achievable by dynamo amplification~($B_{\rm max}$, red curve)
as a function of the coherence length $l$.
The magnetic energy becomes equal to the turbulent kinetic energy on the blue dashed line~($B_{\rm eq}$).
Field fluctuations are cut off by the viscous dissipation below $l_\nu$~(vertical black solid line)
and by the Ohmic loss below $l_{\rm Ohm}$~(vertical black dashed line), respectively.
At $l > {\rm max}[l_\nu,l_{\rm Ohm}] $, fields above $B_{\rm diss}$~(green dashed line) dissipate dominantly by the ambipolar diffusion.
The panels show the cases with (a) $(\zeta_{\rm CR,0}, n_{\rm H}) = (0, 10\ {\rm cm}^{-3})$, (b) $(0, 10^{10}\ {\rm cm}^{-3})$, (c) $(10^{-17}\ {\rm s}^{-1}, 10\ {\rm cm}^{-3})$, and (d) $(10^{-17}\ {\rm s}^{-1}, 10^{10}\ {\rm cm}^{-3})$, respectively.
Note that in panels (a) and (c), $l_{\rm Ohm}$ is much smaller than $l_\nu$ and is not shown.}
\label{fig:dynamo_Bmax}
\end{center}
\end{figure*}

The degree of field dissipation can be judged from the magnetic Reynolds number Rm defined 
by the ratio of the flux-freezing term~($\nabla \times (\bm{{\rm v}}_{\rm n} \times \bm{B}$)) to the dissipation term~(second term) in Eq. \eqref{eq:diff_eq2}:
\begin{equation}
{\rm Rm}(L_{\rm B}) \equiv \frac{{\rm v}_{\rm n} L_{\rm B}}{\eta_{\rm ambi} + \eta_{\rm Ohm}},
\label{eq:Rm}
\end{equation}
where $L_{\rm B}$ is the coherence length of the field.
From the field dissipation time $t_{\rm dis}(L_{\rm B}) = L_{\rm B}^2/(\eta_{\rm ambi} + \eta_{\rm Ohm})$ and the gas dynamical time $t_{\rm dyn}(L_{\rm B}) = L_{\rm B}/{\rm v}_{\rm n}$, ${\rm Rm}(L_{\rm B})$ is rewritten as
\begin{equation}
{\rm Rm}(L_{\rm B}) = \frac{t_{\rm dis}(L_{\rm B})}{t_{\rm dyn}(L_{\rm B})}.
\label{eq:Rm_2}
\end{equation}
When ${\rm Rm}(L_{\rm B}) < 1$, the magnetic field coherent over $L_{\rm B}$ decouples from the fluid motion and then dissipates.

First, we study the case of a global magnetic field coherent over the entire cloud size~($L_{\rm B} \sim \lambda_{\rm J}$).
This happens when the field lines are dragged by the cloud collapse without turbulent motions at smaller scales. 
In evaluating ${\rm Rm}(\lambda_{\rm J})$, we set $t_{\rm dyn}(\lambda_{\rm J}) \sim t_{\rm col}$ in Eq. \eqref{eq:Rm_2}, i.e., 
the collapse proceeds in the free-fall manner.
In Figure \ref{fig:Rm_rev}, we show the contour map of ${\rm Rm}(\lambda_{\rm J})$ as a function of the magnetic field and density 
for $\zeta_{\rm CR, 0} = 0$ (top) and $10^{-17}\ {\rm s}^{-1}$ (bottom). 
The contours are plotted from $\log {\rm Rm}(\lambda_{\rm J}) = 0$ to 10 with the interval of 2~(black lines).
The thick lines show ${\rm Rm}(\lambda_{\rm J}) =1$, above which the field dissipates from the cloud.
The white lines indicate the critical field strength $B_{\rm crit}$,
\begin{equation}
B_{\rm crit} = \left(\frac{4 \pi G M_{\rm J} \rho}{\lambda_{\rm J}}\right)^{1/2},
\label{eq:B_coll}
\end{equation}
at which the gravity is balanced by the magnetic pressure, and the cloud contraction is prohibited if $B > B_{\rm crit}$.
Figure \ref{fig:Rm_rev} shows ${\rm Rm}(\lambda_{\rm J}) > 1$ for the cloud scale field at all the densities as long as the field is weaker than the critical~($B < B_{\rm crit}$)
and the flux is conserved.
The magnetic diffusivities become smaller due to the CR ionization and the field couples more tightly to the gas at $\lesssim 10^{11}\ {\rm cm}^{-3}$. 
As discussed in Section \ref{subsec:fiducial}, the ionization degree and so ${\rm Rm}(\lambda_{\rm J})$ were underestimated by more than two orders of magnitude 
at $\sim 10^{14}\mbox{-}10^{18}\ {\rm cm}^{-3}$ in the previous calculations~\citep{Maki2004,Maki2007,Susa2015}.  
Even so, their ${\rm Rm}(\lambda_{\rm J})$ was already larger than unity when $B < B_{\rm crit}$ and their conclusion that the field does not 
dissipate remains valid. 

Magnetic fields fluctuating at smaller scales are more subject to dissipatation. 
Turbulence in the cloud may stretch and twist the field lines and generate fields coherent over the eddy scale, 
$L_B \sim l$~\citep[the so-called small scale dynamo; e.g.,][]{Brandenburg2005}.
Here, we also consider the case that turbulence is driven at the cloud scale by the gravitational contraction~\citep{Federrath2011}, 
and turbulent velocity scales with the eddy size following the Kolmogorov law~\citep{Kolmogorov1941}:
\begin{equation}
{\rm v}(l) = {\rm v}_{\rm ff} \left(\frac{l}{\lambda_{\rm J}}\right)^{1/3} \;\;\; ( l_\nu \leq l \leq  \lambda_{\rm J} ),
\label{eq:v_l}
\end{equation}
where ${\rm v}_{\rm ff} \sim \lambda_{\rm J}/t_{\rm col}$ is the free-fall velocity. 
The viscous dissipation scale $l_\nu$ is the scale where the eddy turnover time $t_{\rm eddy}(l) \equiv l/{\rm v}(l)$ 
becomes equal to the viscous dissipation time $t_{\rm vis}(l) \equiv l^2/\nu$:
\begin{equation}
\frac{l_\nu}{\lambda_{\rm J}} = \left(\frac{\nu}{{\rm v}_{\rm ff}\lambda_{\rm J}}\right)^{3/4},
\label{eq:l_nu}
\end{equation}
where $\nu = c_s/(n_{\rm H} \sigma_{\rm nn})$ is the kinematic viscosity, and $\sigma_{\rm nn} = 10^{-16}\ {\rm cm}^2$ 
the collision cross section between neutral particles~\citep{Subramanian1998}.
In this case, the magnetic Reynolds number at a scale $l$ is calculated by substituting $L_B = l$ and  ${\rm v}_{\rm n} = {\rm v}(l)$ in Eq. \eqref{eq:Rm} as
\begin{equation}
{\rm Rm}(l) = \frac{{\rm v}(l)\ l}{\eta_{\rm ambi} + \eta_{\rm Ohm}} \sim \frac{t_{\rm dis}(l)}{t_{\rm eddy}(l)},
\label{eq:Rm_l}
\end{equation}
and magnetic field dissipation occurs when ${\rm Rm}(l) < 1$.

The condition of ${\rm Rm}(l) < 1$ is rewritten from the approximate form for $\eta_{\rm ambi}$ in Eq. \eqref{eq:eta_ambi_app} as
\begin{equation}
\begin{split}
B > B_{\rm diss}(l) &\sim \left(4 \pi \mu_{\rm i,n} <\sigma {\rm v}>_{\rm i,n} n_{\rm n} n_e u_{\rm ff} \lambda_{\rm J}\right)^{1/2} \\
&\times \left[\left(\frac{l}{\lambda_{\rm J}}\right)^{4/3} - \left(\frac{l_{\rm Ohm}}{\lambda_{\rm J}}\right)^{4/3}\right]^{1/2},
\end{split}
\label{eq:B_ambi}
\end{equation}
where the Ohmic dissipation scale
\begin{equation}
l_{\rm Ohm} \equiv \lambda_{\rm J} \left(\frac{\eta_{\rm Ohm}}{u_{\rm ff}\lambda_{\rm J}}\right)^{3/4}.
\label{eq:l_Ohm}
\end{equation}
Field fluctuations below $l_{\rm Ohm}$ is cut off by the Ohmic loss. 
At a larger scale $l > l_{\rm Ohm}$, the fields stronger than $B_{\rm diss}(l)$ dissipate by ambipolar diffusion.

Here, we discuss how much the field is amplified by the small scale dynamo action.
Without dissipation, turbulent eddies with a size $l$ would amplify the field
up to the equipartition level 
\begin{equation}
B_{\rm eq}(l) \sim B_{\rm crit} \left(\frac{l}{\lambda_{\rm J}}\right)^{2/3} \;\;\; (l_\nu \leq l \leq \lambda_{\rm J}),
\label{eq:B_eq}
\end{equation}
where the magnetic energy becomes equal to the turbulent kinetic energy, $B_{\rm eq}(l)^2/(8\pi \rho) \sim {\rm v}(l)^2/2$. 
Further amplification is suppressed due to the back-reaction from the field.
With dissipation taken into account, any small-scale field at $l \leq l_{\rm Ohm}$ is erased by the Ohmic loss, 
while the growth beyond $B_{\rm diss}(l)$ is suppressed by ambipolar diffusion at $l > l_{\rm Ohm}$.
In summary, the maximum magnetic field strength $B_{\rm max}(l)$ achievable by dynamo amplification would be the smaller of $B_{\rm eq}(l)$ and $B_{\rm diss}(l)$:
\begin{equation}
B_{\rm max}(l) = \begin{cases}
0 & ( l \leq {\rm max}[l_\nu, l_{\rm Ohm}] ), \\
{\rm min}[B_{\rm eq}(l), B_{\rm diss}(l)] &  ( l > {\rm max}[l_\nu, l_{\rm Ohm}]).
\end{cases} 
\label{eq:B_max}
\end{equation}
In Figure \ref{fig:dynamo_Bmax}, we show $B_{\rm max}(l)$~(red solid), $B_{\rm eq}(l)$~(blue dashed), and $B_{\rm diss}(l)$~(green dashed) for the cases 
with (a) $(\zeta_{\rm CR,0}, n_{\rm H}) = (0, 10\ {\rm cm}^{-3})$, (b) $(0, 10^{10}\ {\rm cm}^{-3})$, (c) $(10^{-17}\ {\rm s}^{-1}, 10\ {\rm cm}^{-3})$, and (d) $(10^{-17}\ {\rm s}^{-1}, 10^{10}\ {\rm cm}^{-3})$, respectively.
The vertical black solid and dashed lines indicate the viscous~($l_\nu$) and Ohmic dissipation~($l_{\rm Ohm}$) scales, respectively.
In panels (a) and (c), $l_{\rm Ohm}$ is much smaller than $l_\nu$ and is not shown.
Note that for $n_{\rm H} < 10^9\ {\rm cm}^{-3}$~(or $> 10^9\ {\rm cm}^{-3}$), the results are qualitatively the same as those of $n_{\rm H} = 10\ {\rm cm}^{-3}$~(or $= 10^{10}\ {\rm cm}^{-3}$, respectively).

First, we see the case without CR ionization~(Figure \ref{fig:dynamo_Bmax} a and b).
Note that the magnetic field is amplified in the eddy turnover timescale $t_{\rm eddy}(l) \propto l^{2/3}$, 
so grows faster at smaller scales.
When $n_{\rm H} = 10\ {\rm cm}^{-3}$~(Figure \ref{fig:dynamo_Bmax} a), dynamo amplification first occurs near the viscous scale $l \sim l_\nu$ 
until saturation at $B_{\rm diss}(l)$ by ambipolar diffusion.
The magnetic field strength set by the dissipation $B_{\rm diss}(l)$ is more than half the equipartition at that scale.
At a larger scale $l \gtrsim 5 \times 10^{-4}\lambda_{\rm J}$, the field can grow up to the equipartition level without dissipation.
When $n_{\rm H} = 10^{10}\ {\rm cm}^{-3}$~(Figure \ref{fig:dynamo_Bmax} b), since the Ohmic diffusivity is much larger than that at $n_{\rm H} = 10\ {\rm cm}^{-3}$~(Figure \ref{fig:res}), the Ohmic dissipation scale appears above the viscous scale and the fields at $l_\nu < l < l_{\rm Ohm}$ are cut off.
Also at larger scales $l_{\rm Ohm} < l <  0.1 \lambda_{\rm J}$, the magnetic field grows only up to a small fraction of the equipartition level due to ambipolar diffusion.
At later times, once the dynamo starts to operate at near the Jeans scales $l \gtrsim 0.1 \lambda_{\rm J}$, 
the field can reach the equipartition level,  if long enough time is available for amplification.
Qualitatively the same behavior is observed in the presence of CR~(Figure \ref{fig:dynamo_Bmax} c and d).
When $n_{\rm H} = 10\ {\rm cm}^{-3}$~(Figure \ref{fig:dynamo_Bmax} c), since CR ionization strengthens the coupling between the magnetic field and gas, 
field dissipation is suppressed at any scale. 
Even when $n_{\rm H} = 10^{10}\ {\rm cm}^{-3}$~(Figure \ref{fig:dynamo_Bmax} d), the field growth is suppressed by dissipation 
only at $l \gtrsim 10^{-3} \lambda_{\rm J}$, much smaller than in the no CR case.
Therefore, in a turbulent cloud, the dynamo action can potentially amplify the magnetic field to the equipartition level 
from very small scales~($l \sim 10^{-3}-10^{-1} \lambda_{\rm J}$) up to the largest scale~($\lambda_{\rm J}$).
This conclusion is consistent with the previous analytical studies~\citep{Schleicher2010,Schober2012a}.


\subsection{Back-reactions of strong magnetic fields on thermal evolution}
\label{subsec:B_backreact}

\begin{figure*}
\begin{center}
{\includegraphics[scale=1.1]{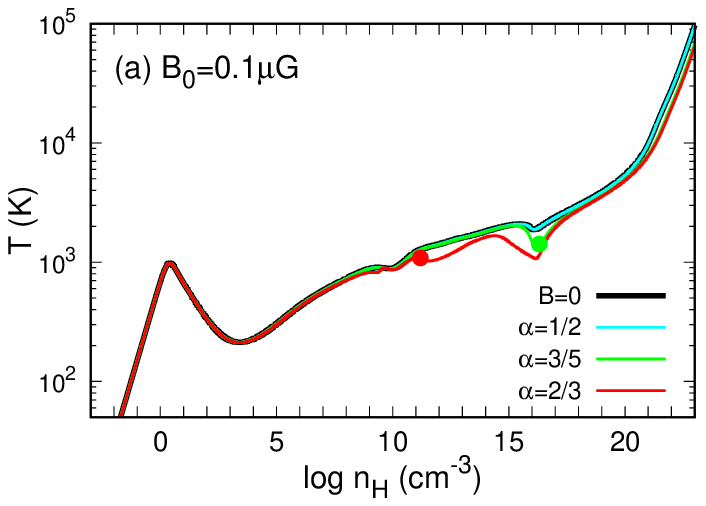}}
{\includegraphics[scale=1.1]{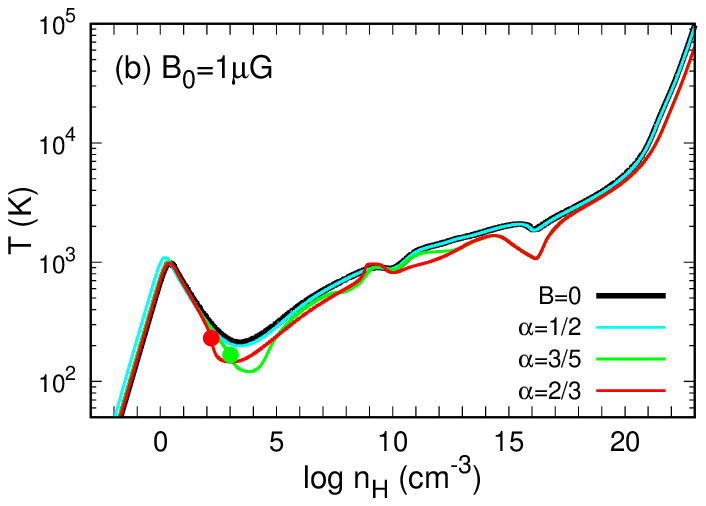}}\\
{\includegraphics[scale=1.1]{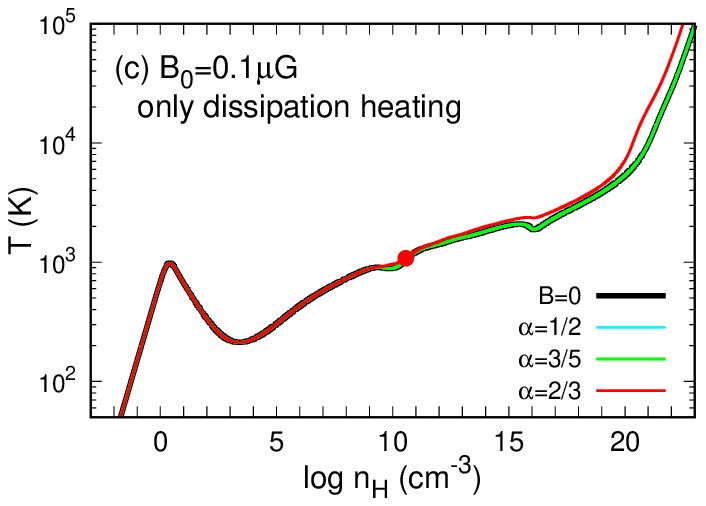}}
{\includegraphics[scale=1.1]{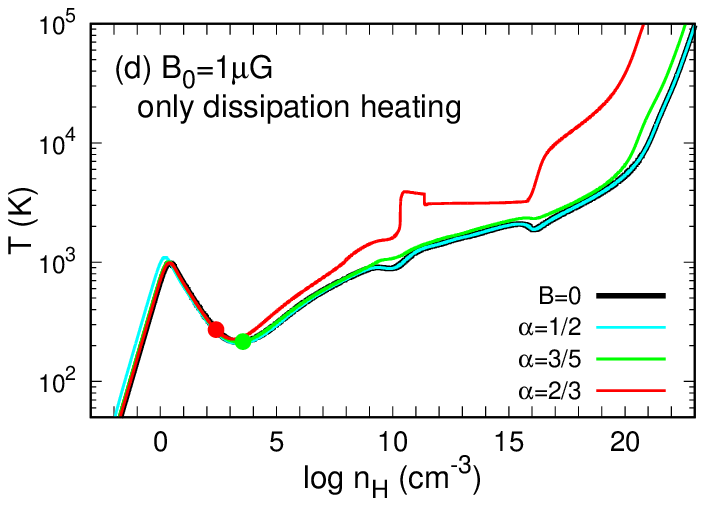}}\\
\caption{Impact of strong magnetic fields on thermal evolution for the magnetic field strengths at $n_{\rm H}=1 {\rm cm}^{-3}$ of $B_0 = 0.1$ and $1\ \mu{\rm G}$.
Panels (a) and (b) show the cases with both the magnetic pressure and heating considered, while panels (c) and (d) show those with the magnetic heating alone.
The black lines show the no-field cases.
Light blue, green and red lines correspond to the cases with $\alpha = 1/2, 3/5$ and $2/3$, respectively.
Filled circles indicate the density where the field reaches $B_{\rm crit}$.
Note that for the cases where the field never reaches $B_{\rm crit}$, the filled circles do not appear.}
\label{fig:mag_nt}
\end{center}
\end{figure*}

So far, we have neglected back-reactions of the magnetic fields, such as the magnetic pressure and 
heating due to the field dissipation under the weak field assumption. 
The fields are amplified with the cloud contraction and those effects can be important. 
\cite{Schleicher2009} showed that heating due to ambipolar diffusion becomes remarkable and the thermal evolutionary 
track deviates from that in the no field case, once the field becomes as strong as $B_{\rm crit}$.
With such strong fields, the collapse would slow down by the magnetic pressure, but this effect has not been 
properly taken into account in their calculation. 
Here, we calculate the thermal evolution by including both of those effects. 

We describe our model for the magnetic pressure and heating in turn. 
The magnetic pressure effect is modeled by replacing $G$ in the free-fall time~(Eq. \ref{eq:t_str}) 
with the effective gravitational constant $G_{\rm eff}$
\begin{equation}
G_{\rm eff} = G \left[1-\left(\frac{B}{B_{\rm crit}}\right)^2\right].
\label{eq:G_eff}
\end{equation}
If we apply this prescription naively, the collapse stops when $B$ reaches $B_{\rm crit}$. 
In reality, however, accretion of the surrounding materials makes the cloud contract further quasi-statically. 
We mimick this effect by setting the upper limit on the collapse time as 
$t_{\rm col, max} = t_{\rm col}/\sqrt{1-f}$, where $f=0.95$~\citep{Omukai2005}.
The dissipation rate of the field energy per unit volume is given by $\bm{J} \cdot \bm{E}$.
In a weakly ionized gas, $\bm{E}$ is related to $\bm{J}$ through the generalized Ohm's law~\citep{Nakano1986}:
\begin{equation}
\bm{E}= \frac{4 \pi}{c^2} \left[\eta_{\rm Ohm} \bm{J} + \eta_{\rm Hall} \bm{J} \times \bm{e}_B - \eta_{\rm ambi} \left(\bm{J} \times \bm{e}_B \right) \times \bm{e}_B\right],
\label{eq:ohm_law}
\end{equation}
so that from Amp\`ere's law,
\begin{equation}
\bm{J} \cdot \bm{E} = \frac{\eta_{\rm Ohm}}{4 \pi} \left|\nabla \times \bm{B} \right|^2 + \frac{\eta_{\rm ambi}}{4 \pi} \left|\left(\nabla \times \bm{B}\right) \times \bm{e}_B\right|^2.
\label{eq:gam_mag}
\end{equation}
Assuming the field coherent over the cloud scale $\sim \lambda_{\rm J}$, the heating rate per unit mass is given by
\begin{equation}
\Gamma_{\rm mag} \equiv \frac{\bm{J} \cdot \bm{E}}{\rho} \sim \frac{\eta_{\rm ambi} + \eta_{\rm Ohm}}{4 \pi \rho} \left(\frac{B}{\lambda_{\rm J}}\right)^2,
\label{eq:gam_mag2}
\end{equation}
which is taken into account in solving the energy equation Eq. \eqref{eq:EoE}.

Next we consider the evolution of magnetic energy density~($E_{\rm B} = B^2/8 \pi$). 
Without dissipation~$\Gamma_{\rm mag} = 0$, the magnetic field is amplified by compression. 
We parametrize this effect as $B \propto \rho^{\alpha}$ and study three cases of $\alpha = 2/3, 3/5$, and  $1/2$.
The first case $\alpha = 2/3$ corresponds to the spherical collapse. 
With strong magnetic field, the collapse will be sheet-like
as the contraction proceeds more along the field. 
In this limit, $\alpha = 1/2$. 
Considering also the field dissipation, the evolution of magnetic energy density is calculated from
\begin{equation}
\frac{dE_{\rm B}}{dt} = \frac{2 \alpha E_{\rm B}}{\rho} \frac{d \rho}{dt} - \rho \Gamma_{\rm mag}.
\label{eq:ene_mag}
\end{equation}

The temperature evolution with the magnetic field effects is presented in Figure \ref{fig:mag_nt}.
Panels (a) and (b) show the cases with both the magnetic pressure and heating taken into account, 
while panels (c) and (d) the cases with the magnetic heating alone, as in \cite{Schleicher2009}.
Magnetic field strength is parametrized by $B_0$, the value at $n_{\rm H}=1\ {\rm cm}^{-3}$, 
and so the value at the initial density $n_{\rm H, ini} = 5.6 \times 10^{-3}\ {\rm cm}^{-3}$ is given by $B_0 n_{\rm H, ini}^\alpha$. 
We study two cases for $B_0$, $B_0 = 0.1$ and $1\ \mu{\rm G}$~(left and right panels, respectively). 
In both panels, the black lines indicate the case with $B = 0$, and 
light blue, green and red lines are for those with $\alpha = 1/2, 3/5$ and $2/3$, respectively.
The filled circles indicate the densities where the field reaches the critical value $B_{\rm crit}$.
Note that the filled circles do not appear for the models where the field is always weaker than $B_{\rm crit}$.

In cases with both the magnetic pressure and heating effects taken into account~(Figure \ref{fig:mag_nt} a and b), 
temperature becomes somewhat lower than in the no field case ($B=0$) after the field reaches the critical value 
$B_{\rm crit}$~(filled circles) since the compressional heating rate is lowered 
due to the delayed collapse by the magnetic pressure. 
With the larger initial field $B_0$, the critical field strength is reached at lower density and temperature deviation is observed 
in a wider density range.
If we include only the field dissipation heating, the effect on the thermal evolution 
appears in the opposite direction (Figure \ref{fig:mag_nt} c and d). 
Now the temperature is higher due to the dissipation heating than in the $B=0$ case.  
This is most conspicuous in the case of $B_0 = 1\ \mu{\rm G}$ and $\alpha = 2/3$~(red curve in panel d), 
where the temperature is already higher than in the other cases at $> 10^4\ {\rm cm}^{-3}$ and 
exhibits rapid increase at $\sim 10^{10}\ {\rm cm}^{-3}$.
Similar behavior has also been reported in \cite{Schleicher2009}, who did not include 
the slowing down of the collapse by the magnetic pressure.  
The fact that those magnetic field effects work in the opposite directions on the thermal evolution
demonstrates subtlety of the magnetic field modeling. 
For accurate treatment, we need to consider both effects self-consistently, 
hopefully by way of numerical MHD simulations.

\section{Summary and Discussion}
\label{sec:summary}
The ionization degree in the gas controls the coupling between the gas and magnetic field, 
thereby affecting the angular momentum transport and efficiency of binary formation 
from star-forming clouds. 
To evaluate the ionization degree in the primordial gas correctly, 
we have developed a new chemical network where all the reactions are reversed
and have calculated thermal evolution in the collapsing pre-stellar clouds both in cases with and without 
external ionization sources represented by cosmic-ray (CR) incidation. 
From the abundances of chemical species obtained by this model, 
we have calculated the magnetic diffusivities by individual field dissipation processes, i.e., 
the ambipolar and Ohmic diffusion and the Hall effect
and have examined the field dissipation condition for fields coherent over the entire cloud size, as well as those 
fluctuating at smaller scales. 
We have also considered the back-reactions of the magnetic field, the magnetic pressure and dissipation heating, on thermal evolution in the strong field cases.
Below, we briefly summarize the results:

\begin{itemize}

\item The negative charge is always carried dominantly by electrons, 
while the positive charge is carried successively by the following cation species~(Figure \ref{fig:fig_chem}b).
The main cation species is H$^+$ until $\sim 10^{10}\ {\rm cm}^{-3}$.
At $\gtrsim 10^{10}\ {\rm cm}^{-3}$, Li$^+$ becomes the main cation species,
and the ionization degree remains constant $\sim 3 \times 10^{-11}$ until $\sim 10^{14}\ {\rm cm}^{-3}$ due to the long recombination time.
At $10^{14}-10^{18}\ {\rm cm}^{-3}$, Li ionization by the thermal photons trapped in the cloud boosts the ionization degree by an order of magnitude.
This process has not been considered in the previous studies which underestimated the ionization degree by two or three orders of magnitude~(Figure \ref{fig:fig_ele}).
At $\gtrsim 10^{18}\ {\rm cm}^{-3}$, H$^+$ becomes the main cation again.

\item In the presence of CR ionization, although the cloud temperature is initially~($\lesssim 1\ {\rm cm}^{-3}$) higher due to the ionization heating than in the case without CRs, 
it soon becomes lower at higher density~($\gtrsim 1\ {\rm cm}^{-3}$) as a result of enhanced 
H$_2$ and HD cooling by the ionization~(Figure \ref{fig:fig_cr_nT}). 
Above $\sim 10^{11}\ {\rm cm}^{-3}$, the CR is completely attenuated and has no effect on thermal evolution. 

\item We have invented the minimal chemical network that reproduces 
the evolution of temperature, as well as ionization degree, in star-forming clouds of the primordial composition
~(Table \ref{tab:chem_react_red}, Figures \ref{fig:nT_red}-\ref{fig:chem_red_CR12}).
This network consists of 36 reactions among 13 species~(H, H$_2$, e$^-$, H$^+$, H$_2^+$, H$_3^+$, H$^-$, D, HD, D$^+$, Li, LiH, 
and Li$^+$) in the absence of CRs, and the additional 10 reactions and 2 species (He and He$^+$) are needed in the presence of CRs. 

\item The dissipation of a global magnetic field at the cloud scale~($\sim$ Jeans scale $\lambda_{\rm J}$) is negligible,
as long as the field is not so strong as to prevent the gravitational collapse~(Figure \ref{fig:Rm_rev}).
Since the magnetic diffusivity is inversely proportional to the ionization degree, 
CR ionization reduces the diffusivity and strengthens the coupling between the gas and the field at $< 10^{11}\ {\rm cm}^{-3}$.

\item Turbulent motions generate magnetic fields fluctuating at smaller scales~($l < \lambda_{\rm J}$).
At $n_{\rm H} = 10\ {\rm cm}^{-3}$, magnetic dissipation is negligible all the way 
from the viscous dissipation scale $l_\nu$ up to the cloud scale $\lambda_{\rm J}$~(Figure \ref{fig:dynamo_Bmax} a, c).
Dynamo amplification then occurs until the magnetic energy becomes equal to the turbulent energy.
At higher density of $n_{\rm H} = 10^{10}\ {\rm cm}^{-3}$, owing to the higher magnetic diffusivity~(Figure \ref{fig:res}), 
dissipation occurs at scales below $l \sim 0.1\ \lambda_{\rm J}$~(Figure \ref{fig:dynamo_Bmax} b).
CR ionization lowers the magnetic diffusivity, making dissipation negligible up to $l \sim 10^{-3}\ \lambda_{\rm J}$, much smaller than in the no CR case~(Figure \ref{fig:dynamo_Bmax} d).
The magnetic fields can potentially be amplified until the equipartition level
at $l \gtrsim 10^{-1}\ \lambda_{\rm J}$ without CR incidation~(or at $l \gtrsim 10^{-3}\ \lambda_{\rm J}$ with CR incidation). 

\item When the magnetic field is so strong as to suppress the collapse, thermal evolution shows deviation from that in the no field case.
In cases with both the magnetic pressure and dissipation heating considered, the compressional heating rate decreases due to the delayed collapse by the magnetic pressure, 
and the temperature becomes lower than in the no field case~(Figure \ref{fig:mag_nt} a, b).
If we include the dissipation heating alone, thermal evolution deviates in the opposite direction, 
i.e., the temperature becomes higher due to the dissipation heating than in the no field case~(Figure \ref{fig:mag_nt} c, d).
Thus, thermal evolution is sensitively affected by the modeling of the magnetic pressure and dissipation heating.

\end{itemize}

Our results in the strong field cases are especially subject to modifications if some of our simple assumptions are relaxed.
For example, we assume the field increases at a constant rate $\alpha$ by the cloud collapse as Eq. \eqref{eq:ene_mag}
and also the field is coherent over the cloud scale~($\sim$ Jeans scale).
The cloud, however, contracts more along the field line due to the magnetic pressure, 
and will become more oblate with contraction. 
This means that the rate of field enhancement $\alpha$ cannot be constant.
In addition, the field coherence length can be larger than the thermal Jeans scale 
in the transverse direction due to the magnetic pressure. 
Those anisotropic structures cannot be captured in the one-zone model and 
multi-dimensional calculations are needed.
\cite{Machida2006,Machida2008} and \cite{Machida2013} performed 
multi-dimensional MHD calculations for the primordial star-formation by 
using the prefixed temperature and ionization-degree evolution as functions of the density
taken from the results in the one-zone model with no-field case.
When magnetic field is so strong as to suppress the collapse, however, 
thermal evolution will clearly deviate from that in the no-field case owing to the magnetic pressure and heating.
This also leads to the deviation of ionization degree and magnetic diffusivity.
For self-consistent treatment, non-ideal MHD equations should be solved by taking the cooling processes, 
magnetic dissipation heating, and chemical reactions into account.

In the weak field cases~($B < 10^{-8}\ {\rm G}$ at $n_{\rm H} = 1\ {\rm cm}^{-3}$),
both the magnetic pressure and dissipation heating do not change the thermal evolution and ionization degree.
Since magnetic dissipation is negligible for a global field,
\cite{Machida2006,Machida2008}'s ideal MHD calculations remain valid,
indicating that for $B > 10^{-11}\ {\rm G}$ at $\sim 1\ {\rm cm}^{-3}$, MHD outflows eject $\sim$ 10\% of the materials from the cloud.
On the other hand, \cite{Machida2013} found by resistive MHD calculations 
that for $B > 10^{-12}\ {\rm G}$ at $\sim 1\ {\rm cm}^{-3}$, the formation of binary and multiple-star systems is suppressed by magnetic braking.
Adopting the incorrect ionization degree by \cite{Maki2004,Maki2007} at $\sim 10^{14}\mbox{-}10^{18}\ {\rm cm}^{-3}$,
they also found no dissipation for a global field.
Since we have found stronger coupling between the gas and the magnetic field, \cite{Machida2013}'s results remain valid in the weak field cases.

Although the Hall effect, which has been neglected so far, does not dissipate the field energy~(Section \ref{subsec:resistivity}), 
it can modify the angular momentum distribution in the cloud.
In context of the present-day star formation, \cite{Tsukamoto2015, Tsukamoto2017} studied 
the collapse of rotating dense cores and formation of circumstellar discs by 3D dissipative MHD simulations with the Hall effect
and found that when the angular momentum is parallel to the field, the angular momentum transfer via the magnetic braking is accelerated by the Hall effect, thereby 
suppressing the disc formation.
Conversely, when the angular momentum is anti-parallel to the field, the magnetic braking becomes less efficient, 
leading to the formation of a more massive disc.
Such effects could also be important in the case of primordial star-formation, and would be worth studying in the future. 

\begin{figure}
\begin{center}
{\includegraphics[scale=1.1]{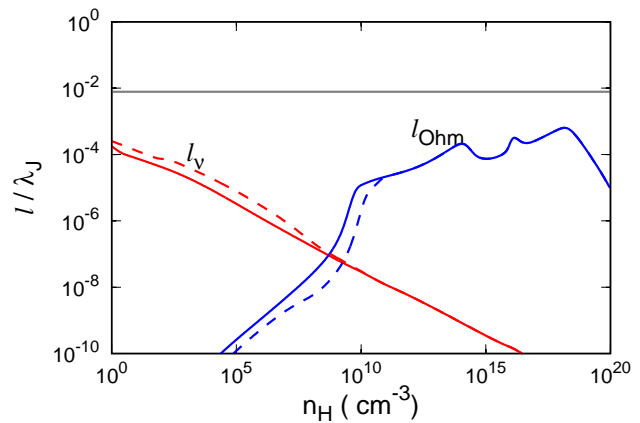}}
\caption{Viscous~(red, $l_\nu$) and Ohmic~(blue, $l_{\rm Ohm}$) dissipation scales as functions of the number density.
The grey line indicates the scale resolved by the highest resolution run in the numerical studies.
Solid and dashed lines correspond to the cases with $\zeta_{\rm CR,0} = 0$ and $10^{-17}\ {\rm s}^{-1}$, respectively.}
\label{fig:dynamo_scale}
\end{center}
\end{figure}

So far, several authors have studied the dynamo action in a collapsing primordial cloud by 3D MHD simulations~\citep{Sur2010,Sur2012,Federrath2011,Turk2012}.
They found that although the dynamo action can be captured by resolving the Jeans scale with more than 32 cells, the growth rate does not converge and increases with the resolution.
\cite{Turk2012} attributed this to insufficient numerical resolution to resolve the smallest driving scale of dynamo,
which corresponds to the larger of the viscous $l_\nu$ and Ohmic dissipation scale $l_{\rm Ohm}$.
In Figure \ref{fig:dynamo_scale}, we compare the viscous~(red, $l_\nu$) and Ohmic~(blue, $l_{\rm Ohm}$) dissipation scales with the most resolved scale in the numerical studies~\citep[grey, 128 cells per Jeans length;][]{Sur2010,Sur2012,Federrath2011,Turk2012}.
Solid and dashed lines correspond to the cases with $\zeta_{\rm CR,0} = 0$ and $10^{-17}\ {\rm s}^{-1}$, respectively.
This figure shows that the smallest driving scale lies $10^{-5}-10^{-1}$ times below the numerically resolved scale, 
which needs to be resolved to obtain the reliable dynamo growth rate in the primordial gas.

\begin{figure}
\begin{center}
{\includegraphics[scale=1.1]{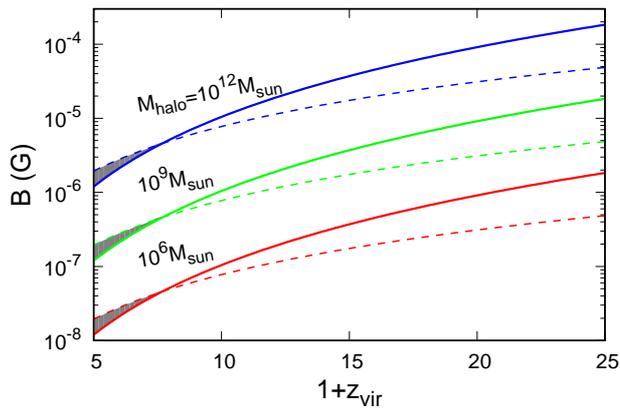}}\\ 
\caption{Condition for the charge separation during the halo virialization.
Colors show the cases with $M_{\rm halo} = 10^6\ {\rm M}_\odot$~(red), $10^9\ {\rm M}_\odot$~(green), and $10^{12}\ {\rm M}_\odot$~(blue), respectively.
Above the solid line, charged species are removed from the halo through ambipolar diffusion.
Below the dashed line, the gravity overcomes the magnetic pressure and the halo can collapse.
Star formation from the charge separated gas occurs only in the grey-shaded regions.}
\label{fig:Li_problem}
\end{center}
\end{figure}

\ 

Finally, we discuss possible roles of the magnetic fields in solving 
so-called the Li problem in the Big Bang Nucleosynthesis (BBN), which has recently been studied by \cite{Kusakabe2015,Kusakabe2019}.
The $^7$Li abundance of observed Galactic metal-poor stars exhibits 
the universal value, so-called the Spite plateau, 
for those in the metallicity range of [Fe/H] $\sim$ -3...-2~\citep{Spite1982}.
This $^7$Li plateau is, however, three times smaller than the standard BBN prediction. 
The unknown origin of this deviation is called the Li problem.
\cite{Kusakabe2015,Kusakabe2019} insisted that if the primordial gas in a halo is strongly magnetized already at the turnaround epoch, 
charged species including $^7$Li$^+$ drift out of the halo via ambipolar diffusion during the collapse and 
the stars formed there will show the $^7$Li depletion from the initial BBN value.
We re-examine the condition for this charge separation to occur by using our results. 

Consider a halo with mass $M_{\rm halo}$ virializing at the redshift $z_{\rm vir}$.
Starting from the turnaround, it takes the free-fall time $t_{\rm col, turn}$ for the virialization.
On the other hand, charged species drift out of the halo in the ambipolar diffusion timescale, 
$R_{\rm halo}^2/ \eta_{\rm ambi}$, where $R_{\rm halo}$ is the halo radius at the turnaround. 
The condition for the charged species to be removed from the halo is 
$t_{\rm col, turn} > R_{\rm halo}^2/\eta_{\rm ambi}$, which can be rewritten with respect to the magnetic field as
\begin{equation}
B > 10^{-7} \left(\frac{y(e)}{10^{-4}}\right)^{1/2}
\left(\frac{M_{\rm halo}}{10^6\ {\rm M}_\odot}\right)^{1/3} \left(\frac{1+z_{\rm vir}}{10}\right)^{11/4} {\rm G}, 
\label{eq:cond1}
\end{equation}
where we have used the ambipolar diffusivity $\eta_{\rm ambi}$ calculated by substituting 
$({\rm i, n}) = ({\rm H}^+, {\rm H})$ and the typical ionization degree of the IGM $y(e) =10^{-4}$ 
into Eq. \eqref{eq:eta_ambi_app} and 
the halo radius at the turnaround
\begin{equation}
R_{\rm halo} \simeq 700\ \left(\frac{M_{\rm halo}}{10^6\ {\rm M}_\odot}\right)^{1/3} \left(\frac{1+z_{\rm vir}}{10}\right)^{-1}\ {\rm pc}.
\label{eq:R_halo}
\end{equation}
On the other hand, the gas in the halo can collapse only if the gravity overcomes the magnetic pressure:
\begin{equation}
\frac{B^2}{4 \pi R_{\rm halo}} < \frac{G M_{\rm halo}}{R_{\rm halo}^2} \rho_{\rm turn},
\label{eq:cond_collapse}
\end{equation}
where $\rho_{\rm turn}$ is the baryon density at the turnaround. 
This condition is rewritten as
\begin{equation}
B < 8 \times 10^{-8}\ \left(\frac{M_{\rm halo}}{10^6\ {\rm M}_\odot}\right)^{1/3} \left(\frac{1+z_{\rm vir}}{10}\right)^2\ {\rm G}.
\label{eq:cond2}
\end{equation}
Star formation from the charge separated gas occurs only if the magnetic field satisfies both conditions of
Eqs. \eqref{eq:cond1} and \eqref{eq:cond2}.
This range of magnetic fields where the charge separation occurs is shown by the grey-shaded regions in Figure \ref{fig:Li_problem}
for the halos with masses $M_{\rm halo} = 10^6\ {\rm M}_\odot$~(red), $10^9\ {\rm M}_\odot$~(green), and $10^{12}\ {\rm M}_\odot$~(blue, respectively).
We can see that the charge separation occurs only in the very narrow ranges of the redshift and magnetic field, 
which, in addition, will completely disappear if the ionization degree in the IGM is higher than $\sim 3 \times 10^{-4}$~(see Eq. \ref{eq:cond1}).
Since the IGM reionization is almost over at $z_{\rm vir} \lesssim 7$ and 
the ionization degree then is likely to be higher than this value, 
we expect that the charge separation did not occur in the Milky Way building block halos. 

\section*{Acknowledgments}
We would like to thank Francesco Palla for inspire us to study the Li$^{+}$ charge separation in the metal-poor gas.
We also thank Masahiro N. Machida, Yusuke Tsukamoto and Jiro Shimoda for their helpful comments on 
the magnetic dissipation and dynamo amplification.
This work is supported in part by MEXT/JSPS KAKENHI grants (DN:16J02951, KO:25287040, 17H01102, 17H02869, HS: 17H01101, 17H02869).




\bibliographystyle{mnras}

\label{lastpage}
\end{document}